\documentclass[preprint,12pt,authoryear]{elsarticle}
\usepackage{hyperref}
\usepackage{graphicx}%
\usepackage{subfigure}
\usepackage{multirow}%
\usepackage{amsmath,amssymb,amsfonts,amsthm}%
\usepackage{bm}
\usepackage{mathrsfs}%
\usepackage[title]{appendix}%
\usepackage{xcolor}%
\usepackage{textcomp}%
\usepackage{manyfoot}%
\usepackage{booktabs}%
\usepackage{algorithm}%
\usepackage{algorithmicx}%
\usepackage{algpseudocode}%
\usepackage{listings}%
\usepackage{float}
\usepackage{caption}
\usepackage{subcaption}
\usepackage{array}
\usepackage{rotating}
\usepackage{color,soul}

\begin{document}
\begin{frontmatter}
\title{\textbf{A Two-Phase Model of Early Atherosclerotic Plaque Development with LDL Toxicity Effects}}
\author[inst1]{Abdush Salam Pramanik}
\author[inst1]{Bibaswan Dey\corref{mycorrespondingauthor}}
\cortext[mycorrespondingauthor]{Corresponding author}
\ead{bibaswandey@nbu.ac.in}
\author[inst2]{G. P. Raja Sekhar}
\address[inst1]{Department of Mathematics, University of North Bengal, Raja Rammohunpur, Darjeeling-734013, West Bengal, India}
\address[inst2]{Department of Mathematics, Indian Institute of Technology Kharagpur, Kharagpur-721302, West Bengal, India}
\begin{abstract}
Atherosclerosis is a chronic inflammatory cardiovascular disease in which fatty plaque is built inside an artery wall. Early atherosclerotic plaque development is typically characterized by inflammatory tissues primarily consisting of foam cells and macrophages. We present a biphasic model that explores early plaque growth to emphasize the role of cytokines (particularly, Monocyte Chemoattractant Protein-1) and oxidized low-density lipoprotein (oxLDL) in monocyte recruitment and foam cell production, respectively. The plaque boundary is assumed to move at the same speed as the inflammatory tissues close to the periphery. This study discusses the oxLDL cholesterols recruitment inside intima and their internalization by the inflammatory cells. Excessive intracellular cholesterol accumulation becomes toxic to macrophage foam cells, leading to cell death beyond a threshold. Our findings reveal that initially, the plaque evolves rapidly, and the growth rate eventually reduces because of the cholesterol-induced toxicity. The present study manifests that higher oxLDL cholesterol flux reduces plaque growth rate, while elevated cytokines flux promotes the corresponding plaque growth behaviour. Between oxLDL cholesterol and intracellular cholesterol, the second one is much more effective towards the growth of inflammatory tissue. The cholesterol-based toxicity-induced cell death parameters are crucial in flattening the plaque growth profile. A detailed analysis of the model presented in this article provides critical insights into the various biochemical and cellular mechanisms behind early plaque development.
\end{abstract}
\end{frontmatter}
\section{Introduction}\label{intro}
\begin{figure}[t]
\centering
\subfigure[]{\label{cross}\includegraphics[width=0.35\textwidth]{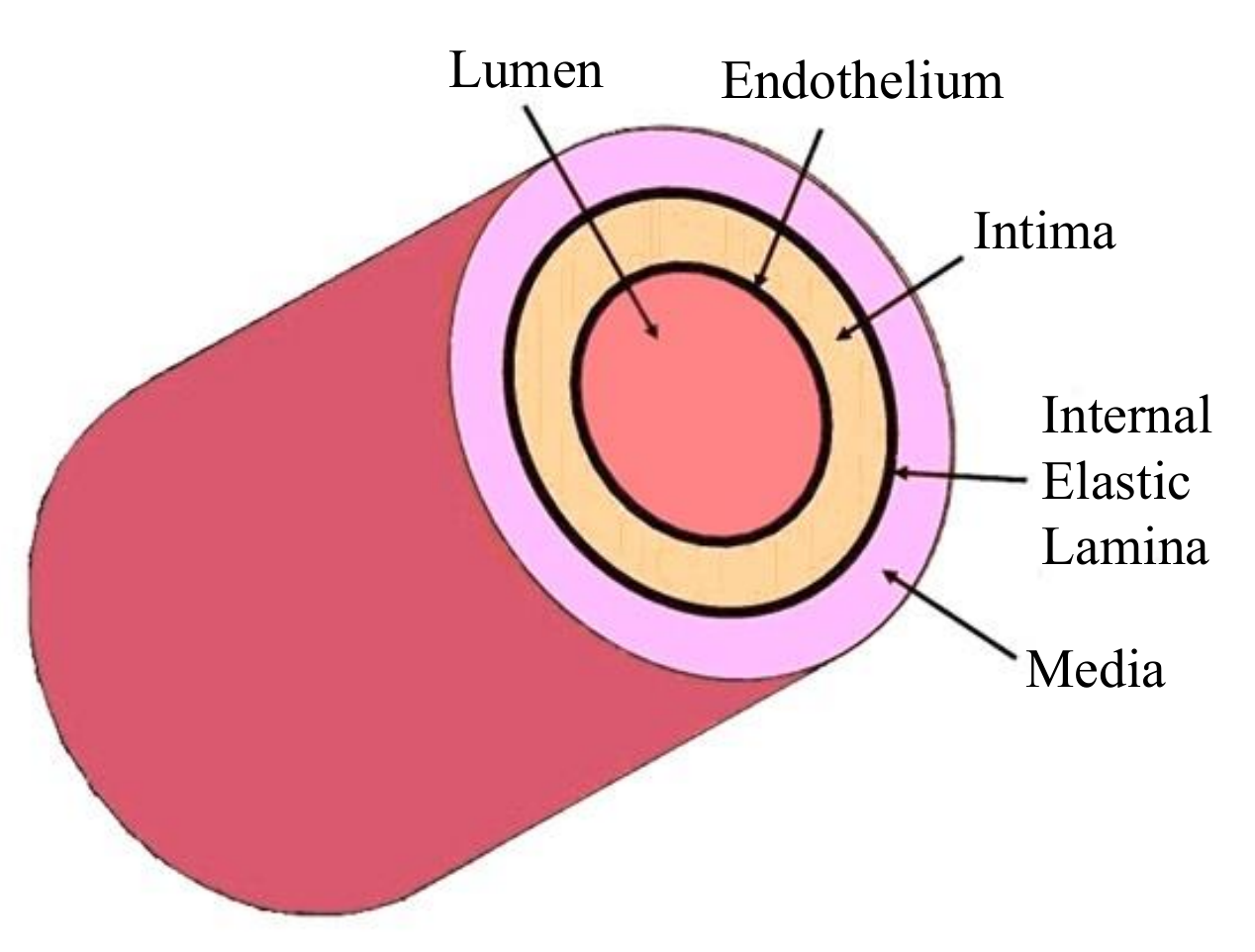}}
\subfigure[]{\label{athero}\includegraphics[width=0.60\textwidth]{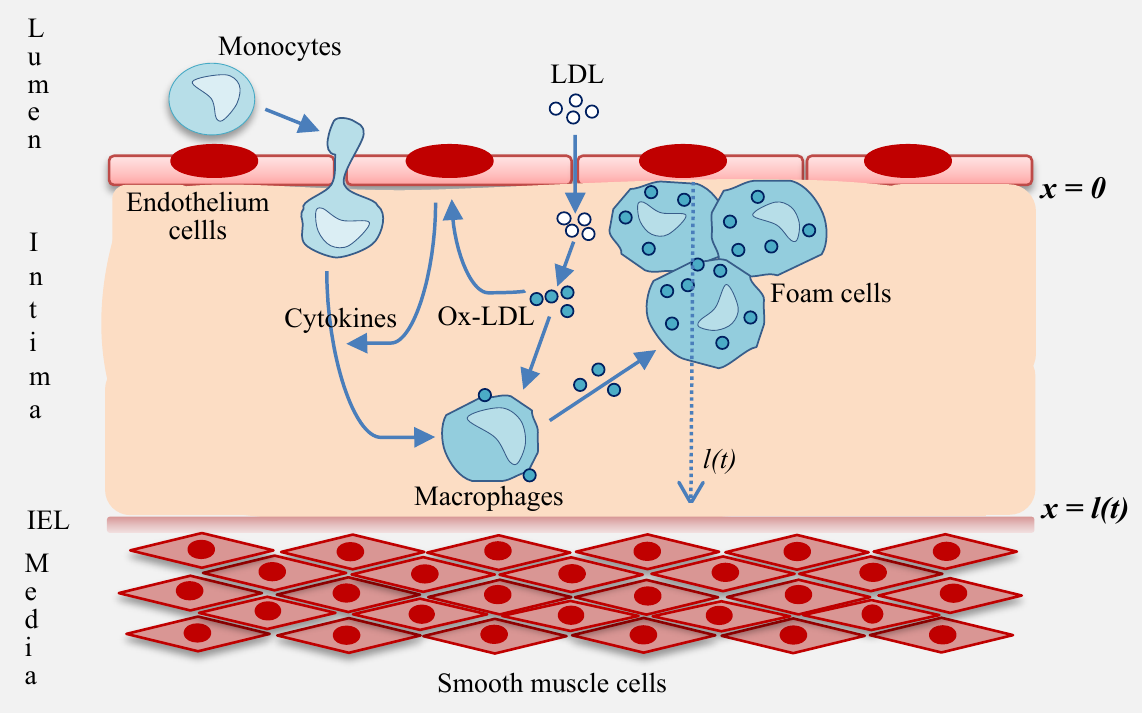}}
\caption{(a) Cross-section of a healthy artery showing the artery wall consists of different cell layers. The innermost layer of the wall is comprised of a thin sheet of endothelium cells, called \emph{the endothelium}. The endothelium forms an interface between the blood flow in the lumen and the rest of the artery wall. Immediately beneath the endothelium, a thin tissue layer lies called the \emph{intima}. The underlying tissue layer is said to be \emph{media} which is separated from the intima by a dense tissue membrane known as \emph{internal elastic lamina (IEL)}. (b) Schematic diagram of the key processes in plaque formation at the early stage. Low-density lipoprotein (LDL), which is present in the bloodstream, enters the intima through the damaged endothelium and undergoes modification to become oxidized LDL (oxLDL). OxLDL induces the secretion of several cytokines from the endothelium cells, including monocyte chemotactic protein-1 (MCP-1), monocyte colony-stimulating factor (M-CSF), and ES-cytokines. In response to the cytokines, particularly for MCP-1, monocytes adhere to the intima from the bloodstream and subsequently differentiate into macrophages. Within the intima, macrophages consume oxLDL to form large lipid-laden foam cells. The accumulation of foam cells produced by macrophages constitutes an early atherosclerotic plaque.}
\end{figure}

\noindent
Atherosclerosis, a type of cardiovascular diseases (CVDs), is an inflammatory disease associated with an excess concentration of low-density lipoprotein (LDL) particles in the bloodstream. The atherosclerotic plaque formation begins with the deposition of sizeable lipid-laden foam cells derived from monocytes which ingest oxidized low-density lipoprotein (oxLDL) through phagocytosis present within intima \citep{libby2002inflammation}. Plaque formation is almost asymptomatic in a patient with atherosclerosis unless diagnosed clinically \citep{davies1993atherosclerosis}. Hence it is essential to detect the plaque at an early stage in order to prevent fatal health complications. Therefore, a complete understanding of plaque growth in early atherosclerosis is crucial. Fig. \ref{cross} depicts the layered structure of a healthy artery wall. When there is turbulence of recirculation in the blood flow into the lumen region, the endothelium gets damaged, particularly in areas of low shear stress. Several factors, such as hypertension, excessive smoking, consuming to much processed foods and chronic alcoholism etc., highly expedite endothelium damage \citep{channon2006endothelium,furchgott1999endothelium}.\\

\noindent
Low-density lipoprotein (LDL) particles from the bloodstream penetrate the artery wall to enter the intima through the injured endothelium, where free oxygen radicals (including superoxide ions) released from endothelial cells modify LDL particles to become oxidized LDL (oxLDL) cholesterol \citep{lusis2000,cobbold2002lipoprotein}. OxLDL cholesterol within intima enables endothelium cells to secrete monocyte chemotactic protein-1 (MCP-1). In addition, endothelial cells express leukocyte adhesion molecules such as vascular cell adhesion molecule-1 (VCAM-1) and intracellular adhesion molecules (ICAM-1), facilitating monocyte attachment on the surface of the endothelial cells from the bloodstream \citep{lusis2000,libby2002inflammation,hansson2006immune}. The monocytes enter the intima in response to chemoattractant gradients like MCP-1 and other cytokines \citep{han2004crp,hansson2006immune,lusis2000}. Subsequently, monocytes differentiate into macrophages inside the intima in response to macrophage colony-stimulating factor (M-CSF) \citep{ross1999atherosclerosis,libby2002inflammation}. Macrophages within the intima consume oxLDL, which causes them to become large, lipid-laden foam cells \citep{lusis2000,calvez2009mathematical,gui2012diverse,hao2014ldl}. Consequently, the foam cells produced along with macrophages form an early plaque in atherosclerosis \citep{chalmers2015bifurcation}. The recruitment of monocytes, their differentiation into macrophages, and subsequent conversion into foam cells have been studied using simplified mathematical models \citep{calvez2009mathematical,cohen2014athero,chalmers2015bifurcation}. The model presented by \citet{ougrinovskaia2010ode} assumes Michaelis-Menten kinetics for the uptake of oxidized LDL (oxLDL) by macrophages, with the recruitment of macrophages dependent on oxLDL levels. Additionally, they utilized a sigmoidal function to represent the saturation of the uptake rate. The models of \citet{el2007atherosclerosis,el2009mathematical,el2012reaction} on early atherosclerosis use nonlinear functions at the endothelium boundary to describe the drawing of monocytes or macrophages across the endothelium by accumulated OxLDL within the intima. At this early stage (early atherosclerosis), two essential components of the formed plaque are foam cells and macrophages. These foam cells further secrete more chemoattractants and cytokines to attract monocytes into the intima \citep{chalmers2017nonlinear,ross1999atherosclerosis}. More extensive models proposed by \citet{calvez2009mathematical,calvez2010mathematical} on early atherosclerosis include a cytokine that promotes monocyte recruitment through the endothelium in response to oxLDL consumption by macrophages (foam cells formation).\\

\noindent
Although evidence on the death of macrophages and foam cells is not excessively available in the literature, some studies describe the possible cause. The toxic effect of cholesterol, mainly oxidized LDL, plays a significant role in macrophage or foam cell death. According to \citet{marchant1996oxidation,gotoh1993inhibition}, the toxic environment induced by oxLDL causes macrophages to die, starting after an initial lag phase, as demonstrated \emph{in-vitro} for human monocyte-macrophages. However, this process should be emphasized in detail. LDL particles comprise a protein scaffold (Apolipoprotein B or ApoB) that carries cholesterol, triglycerides, and other lipids \mbox{\citep{schade2017evidence}}. Cholesterol is the predominant lipid component in LDL particles and is the traditional biological marker for LDL \mbox{\citep{qiao2022low}}. Under normal lipemic conditions, LDL cholesterol carries approximately 80\% of the circulating cholesterol \mbox{\citep{acuna2020lipoprotein}}. However, when a macrophage ingests an oxidized LDL particle, it strips the cholesterol from the protein scaffold and stores it as either esterified or unesterified cholesterol \mbox{\citep{chistiakov2016macrophage}}. Consequently, upon ingestion, extracellular oxLDL is internalized into the intracellular cholesterol \mbox{\citep{ahmed2024hdl}}. This excessive cholesterol loading overwhelms the endoplasmic reticulum (ER), inducing ER stress, which promotes apoptosis or other forms of cell death \citep{feng2003endoplasmic,devries2005cholesterol,zhang2003unfolding}. Thus, excessive intracellular cholesterol becomes toxic to inflammatory cells (macrophage foam cells) and inevitably leads to cell death \citep{tabas2002consequences,watson2023lipid}. Furthermore, based on \emph{in-vitro} evidence of mulberry extracts on LDL oxidation and foam cell formation, \mbox{\citet{liu2008mulberry}} demonstrated that macrophages could transform into foam cells when taking up low levels of oxLDL. However, when macrophages ingest large amounts of oxLDL, they lose their ability to process it effectively, leading to cell death. Consequently, a threshold intracellular oxLDL cholesterol concentration exists, beyond which inflammatory cells start to die. The dead foam cells within plaque release free intracellular cholesterol and other cellular debris \mbox{\citep{berliner1995atherosclerosis}}. \\

\noindent
Fig. \mbox{\ref{athero}} outlines the plaque formation at an early stage in a schematic diagram. In the later stages, smooth muscle cells (SMCs) migrate from the media to the intima through the IEL, induced by growth factors like platelet-derived growth factor (PDGF). Stimulated by transforming growth factor-beta (TGF-$\beta$), SMCs produce collagen, with both migrated SMCs and collagen contributing to the formation of the fibrous cap near the endothelium \citep{newby1999fibrous,hansson2006inflammation}. \mbox{\citet{watson2018two}} employ the mixture theory approach to model early fibrous cap formation and smooth muscle cells (SMCs) migration corresponding to the advanced stage of atherosclerosis plaque formation. In another study, \mbox{\citet{watson2020multiphase}} model an advanced plaque tissue, a mixture of three distinct phases: SMC phase, collagen-rich fibrous extracellular matrix (ECM) and a generic tissue phase comprising the remaining plaque constituents. Their study considers collagen cap formation by SMCs in response to diffusible growth factors like PDGF and TGF-$\beta$ from the endothelium. These studies consider the plaque material as advanced and stable without apprising its overall growth. Consequently, the interface between the intima and media remains static. Many researchers believe that atherosclerosis is an obvious application of free boundary models, but for various reasons, most ignore it. In this context, the study of \mbox{\citet{ahmed2022macrophage,ahmed2024hdl}} introduce a free boundary multiphase model corresponding to early atherosclerotic plaques to investigate the effects of impaired macrophage anti-inflammatory behaviour on plaque structure and its growth. Motivated by the above developments of the atherosclerosis model, one can assume that the plaque boundary moves with a velocity equal to the speed of inflammatory tissue. Furthermore, atherosclerosis is a dynamical process of plaque growth that can be modelled using the multiphase mixture theory \citep{breward2002role,byrne2003two,hubbard2013multiphase,dey2016hydrodynamics,dey2018vivo,alam2019mathematical,pramanik2023two}. Multiphase models can produce a detailed description of plaque formation dynamics since they provide a natural framework for volume exclusion and mechanical interaction between plaque constituents. Accordingly, in this study, we present a two-phase model of early atherosclerotic plaque development, characterized by a free plaque boundary. Our primary focus is to investigate the impact of intracellular cholesterol-induced toxicity on plaque development, with emphasizing the role of f-cytokines (particularly MCP-1) and oxLDL in monocyte recruitment and foam cell production, respectively. In the next section, we formulate a mathematical model based on the mixture theory, assuming atherosclerotic plaque as a two-phase material composed of inflammatory and non-inflammatory tissues. The governing equations are nonlinear due to the underlying physics; however, the Michaelis–Menten general kinetics corresponding to the oxLDL metabolism by macrophages can be considered. In order to linearize the oxLDL transport equation, a perturbation expansion method is deployed by assuming the weak nonlinear nature of the oxLDL metabolism. Subsequently, the explicit finite difference method is applied to the resulting leading and first-order equations satisfying the CFL condition.
\section{Mathematical Formulation}
\noindent
While modelling the growth of an early-stage atherosclerotic plaque, we hypothesize that the accumulation of macrophages and foam cells is responsible for the plaque growth. Indeed, the conversion of monocytes into macrophages and rapid oxidization of LDL particles inside the intima take place on a much shorter time scale than the growth of the plaque \mbox{\citep{chalmers2015bifurcation}}. Tracking monocytes within the intima at any given moment would be challenging. Therefore, it is sufficient to view a monocyte as a macrophage and replace LDL with oxLDL. At this point, the plaque is comprised of two separate phases. One comprises macrophages and foam cells, while the other includes interstitial fluid, oxLDL, cytokines, and more. The initial stage of atherosclerosis can be referred to as the inflammatory cell phase, as it predominantly involves macrophages and foam cells derived from macrophages in the development of the inflammatory disease. Therefore, the second phase is non-inflammatory. Inflammatory cells such as monocytes or macrophages migrate and infiltrate based on the regulation of \emph{MCP-1}. This chemokine is activated by the endothelial cells and is a key player in the process. \emph{MCP-1} and other cytokines cause the inflammatory cell  migration, also known as Endothelial Stimulating cytokines or ES cytokines. These cytokines can be considered as single species with a small volume, termed as functional cytokines or f-cytokines. Next, referring to the second paragraph of section \mbox{\ref{intro}}, LDL cholesterol consists of small blood-borne particles that enter the intima from the damaged endothelium and subsequently oxidise to oxLDL cholesterol particles, which are later uptaken by inflammatory cells.
Once again, referring to the third paragraph of section \mbox{\ref{intro}}, when this oxLDL cholesterol goes inside an inflammatory cell, it becomes an intracellular cholesterol. Excessive accumulation of intracellular cholesterol becomes toxic to the cells. Consequently, one must consider oxLDL cholesterol and intracellular cholesterol as two different species similar to f-cytokines.
All three species are driven by diffusion, with oxLDL cholesterol and f-cytokines within the non-inflammatory phase and intracellular cholesterol in the inflammatory cell phase part of the plaque. Their motion can be described using standard transport reaction equations.\\

\noindent
We assume that the domain of the plaque region is $\Omega(t)$ in $\mathbb{R}^d$ ($d=1,2,3$) with moving boundary $\Gamma(t)$ that may be due to the several subparts. To begin, we will establish $\Gamma_1$ as the portion of the plaque boundary due  to the interface between the lumen and intima, or the endothelium. Then, we will define $\Gamma_2(t)$ as the boundary of $\Omega(t)$ that corresponds to the IEL. This boundary separates the intima from the media and increases in size over time due to the accumulation of macrophages and foam cells. Early plaque formation leads to compensatory expansion of an artery, during which the artery extends radially outward while preserving the lumen \citep{korshunov2007vascular}. Consequently, the intima-IEL boundary $\Gamma_2(t)$ can be considered varying with time. We introduce a finite index set $\mathcal{C}$ containing the constituents of the plaque region. Based on our assumptions, the plaque is made up of two types of phases: inflammatory cell ($f$) and non-inflammatory tissue phase ($n$), which we can denote as the set $\mathcal{C}=\{f, n\}$. Alternatively, we can represent it as $\mathcal{C}=\{m, f, e\}$, where $m$ represents monocyte or macrophage, $f$ stands for foam cell, and $e$ denotes extracellular fluid matrix in the plaque region. For this study, we will focus on the two main constituents of the plaque: inflammatory and non-inflammatory tissues, i.e. $\mathcal{C}=\{f, n\}$. However, it is also possible to consider a multiphase model.
\subsection{\textbf{Mass Balance Equations}}
\noindent
For $\alpha\in\mathcal{C}$,  we define $\varphi^{\alpha}$ and $\mathbf{v}^{\alpha}$ respectively as the volume fraction and velocity of each of the phases. The corresponding stress experienced by the $\alpha$-th phase is denoted by $\mathbb{T}^\alpha $. We assume that each phase has the same constant density. Mass balance equations corresponding to each phase can be expressed in the form
\begin{equation}\label{M1}
\frac{\partial \varphi^{\alpha}}{\partial t}+\nabla.\left(\varphi^{\alpha}\mathbf{v}^{\alpha}\right)=\daleth^{\alpha},~~~~ \left(\alpha\in\mathcal{C}\right)
\end{equation}
where $\mathbf{v}^\alpha \in \mathbb{R}^d$ and $\daleth^\alpha \in \mathbb{R}$ is the net source term corresponding to the $i$-th phase. Since the plaque is formed by the constituent of $\mathcal{C}$ and there are no voids, thus
\begin{equation}\label{M2}
\sum_{\alpha\in\mathcal{C}}\varphi^{\alpha} = 1.
\end{equation}
The plaque growth model suggests that specific cells from the bloodstream, such as monocytes or macrophages, can enter the intima (the innermost layer of our arteries) through damaged parts of the endothelium (the inner lining of our blood vessels). These cells consume oxLDL cholesterol and then transform into foam cells. Both macrophages and foam cells are involved in the inflammatory phase of plaque growth. Excessive intracellular cholesterol becomes toxic to inflammatory cells, leading to their death. The resulting dead materials then becomes part of the non-inflammatory phase. We assume that there is no local source or sink of material, and mass is transferred between these phases. Inflammatory cells take up material (oxLDL cholesterol) from the non-inflammatory tissue phase and revert to the non-inflammatory phase upon death. Therefore, the model is closed for mass and consequently satisfies the following:
\begin{equation}\label{M3}
\sum_{\alpha\in\mathcal{C}} \daleth^{\alpha} = 0.
\end{equation}
\subsection{\textbf{Momentum Balance Equations}}
\noindent
Under no external force and negligible inertial effect, the momentum balance equations reduce to the balance equations between intraphase and interphase forces for each inflammatory and non-inflammatory phase \citep{byrne2003two,hubbard2013multiphase}:
\begin{equation}\label{M4}
\nabla.\left(\varphi^{\alpha}\mathbb{T}^\alpha\right)+\mathbf{\Pi}^{\alpha} = 0,~~~~ \left(\alpha\in\mathcal{C}\right),
\end{equation}
where $\mathbb{T}^\alpha \in \mathbb{R}^{d\times d}$ and $\mathbf{\Pi}^{\alpha} \in \mathbb{R}^{d}$ is the interphase force exerted on the $\alpha$-th phase by the other phase. We assume that  the interphase force on the $\alpha$-th is due to Darcy equation obeying the following relations (refer \citet{breward2002role,breward2003multiphase,hubbard2013multiphase})
\begin{equation}\label{M7}
\mathbf{\Pi}^{\alpha} = p \mathbb{I} \nabla \varphi^{\alpha} + \sum_{\beta\in\mathcal{C}}^{\beta\neq \alpha} \mathbf{K} \varphi^{\alpha} \varphi^{\beta}\left(\mathbf{v}^{\beta} - \mathbf{v}^{\alpha}\right),
\end{equation}
in which, $p \in \mathbb{R}$ is the interstitial fluid pressure. The second term on the right-hand side in Eq. (\ref{M7}) represents the relative velocity between the phases characterized by a typical drag coefficient $\mathbf{K} \in \mathbb{R}^{d\times d}$.
\subsection{\textbf{Mass Transport Equations}}
\noindent
It is assumed that oxLDL cholesterol and f-cytokines are present in the non-inflammatory phase, while intracellular cholesterol predominates in the inflammatory cell phase. The concentrations of oxLDL cholesterol, f-cytokines and intracellular cholesterol are governed by the quasi-steady-state reaction-diffusion equations of the form
\begin{equation}\label{M8}
 D_r\nabla.\left(\varphi^{n}\nabla c_r \right) = Q_r, ~~~ r=\{1,3\},
\end{equation}
\begin{equation}\label{M8_1}
 D_2\nabla.\left(\varphi^{f}\nabla c_2 \right) = Q_2,
\end{equation}
in which $c_1$, $c_2$ and $c_3$ are concentrations of oxLDL cholesterol, intracellular cholesterol and f-cytokines, respectively. $D_r \in \mathbb{R}$ ($r=1,2,3$) denotes the diffusion coefficient and $Q_r\in \mathbb{R}$ represents the net source or sink of the $r$-th species \mbox{\citep{watson2018two}}. We assume that the sources $Q_1$, $Q_2$, and $Q_3$ follow Michaelis-Menten kinetics. Since the characteristic timescale of diffusion for the above species is significantly shorter than that of plaque growth, we neglect advective transport within the intima and make a quasi-steady-state approximation \mbox{\citep{hubbard2013multiphase,watson2020multiphase}}.
\subsection{\textbf{Stress Components}}
\noindent
We propose the following constitutive relations corresponding to the stresses $\mathbb{T}^{\alpha}$ ($\alpha=n,f$) on the $\alpha$-th constituent present in Eq. (\ref{M4}) \citep{breward2003multiphase,watson2018two},
\begin{equation}\label{M8a}
 \mathbb{T}^{f}=-\left(p+\Sigma\left(\varphi^{f}\right)+\Lambda(c_3)\right)\mathbb{I},
\end{equation}
\begin{equation}\label{M8b}
\mathbb{T}^{n} = - p\mathbb{I},
\end{equation}
where $\Sigma:[0,~1]\rightarrow \mathbb{R}$ is a nonlinear function of the volume fraction of the inflammatory phase $\varphi^{f}$ and $\Lambda: \mathbb{R}\rightarrow \mathbb{R}$ is another nonlinear function of the concentration of the f-cytokines $c_3$; $\mathbb{I}$ is the identity tensor. To simplify matters, we do not consider viscous forces on either phase, treating these phases as inviscid. As the growth process continues, the increased volume of the foam cells induces a cell-cell interaction. Therefore, beyond a specific value of $\varphi^{f}$, the membrane of each cell starts to deform, and the cells experience stress. In Eq. (\ref{M8a}), $\Sigma$ on the right-hand side represents the pressure during the inflammatory phase caused by cell-cell interactions. The migration of macrophages from the lumen to the intima is affected by f-cytokines. Therefore, the concentration of f-cytokines, denoted as $c_3$, controls the stress in the inflammatory phase. This concentration is expressed by the function $\Lambda(c_3)$.
\subsection{\textbf{Initial and Boundary conditions}}
\noindent
The following initial and boundary conditions are required to complete the model.
\subsubsection{Initial Conditions:}
\noindent
Both the inflammatory phase volume fraction and the location of the plaque outer boundary $\Gamma_2(t)$ are initially assigned as:
\begin{equation}\label{M9}
\varphi^f\left(\mathbf{x},0\right) = \varphi^f_0, ~~ \Gamma_2(t=0)=\Gamma^0,~~~~~~~ \text{for}~ \mathbf{x} \in \mathbb{R}^d.
\end{equation}
\subsubsection{Boundary Conditions:}
\noindent
Monocytes migrate into the intima from the lumen in response to f-cytokines, characterizing an influx of macrophages through the endothelium \citep{hansson2006inflammation,chalmers2015bifurcation}. Consequently, on the interface of the endothelium and intima i.e. at the boundary $\Gamma_1$, we suppose a chemotactic flux of inflammatory cells from the lumen to the intima in response to f-cytokines. Therefore, we can set
\begin{equation}\label{M10}
 \left(\varphi^f\mathbf{v}^f\right)\big|_{\Gamma_1} = \bm{\Xi}(c_3),
\end{equation}
where $\bm{\Xi}(c_3)$ is a vector valued function that characterizes the aforementioned flux at the boundary $\Gamma_1$, and its precise form will be prescribed later. \\

\noindent
The influx of LDL from the lumen into the intima occurs through the damaged portion of the endothelium. In the presence of free oxygen radicals inside the intima, LDL oxidize to become oxLDL. The oxidization of LDL takes a much shorter time than plaque growth, as previously assumed. Consequently, both LDL influx and oxidation can be represented as a single process by the oxLDL influx. The flux of oxLDL cholesterol through $\Gamma_1$ is assumed to be constant and given by \citep{chalmers2015bifurcation},
\begin{equation}\label{M11}
\left(\nabla c_1.\hat{\mathbf{n}}\right)\big|_{\Gamma_1} = - \nu_1,
\end{equation}
where $\hat{\mathbf{n}}$ is the outward unit normal to the surface and $\nu_1$ is constant rate of oxLDL influx. The intracellular cholesterol species ($c_2$) include cholesterol derived from ingested oxLDL through internalization, rather than the endogenous cholesterol content of newly formed cells. Therefore, it can be assumed to have zero flux at the endothelial boundary, as given by,
\begin{equation}\label{M11_2}
 \left(\nabla c_2.\hat{\mathbf{n}}\right)\big|_{\Gamma_1} = 0.
\end{equation}
\noindent
At the boundary $\Gamma_1$, the flux of f-cytokines is modelled by
\begin{equation}\label{M12}
 \left(\nabla c_{3}.\hat{\mathbf{n}}\right)\big|_{\Gamma_1} = -\nu_2c_1 - \nu_3c_3,
\end{equation}
where, $\nu_{2}$ and $\nu_{3}$ are influx parameters. Chemokines produced at the endothelium in response to oxLDL constitute a family of chemoattractant cytokines that play a significant role in selectively recruiting monocytes etc. \citep{lusis2000}. \emph{MCP-1}, like chemokines produced at the endothelium, recruit monocytes, and the incident is expressed by the first term on the right-hand side of Eq. (\ref{M12}). ES cytokines enhance the production of \emph{MCP-1}. As mentioned earlier, \emph{MCP-1}, like chemokines and other ES cytokines, is regarded as f-cytokines. Therefore, in this context, it can be asserted that the production of f-cytokines by the endothelium is selfdependent \citep{chalmers2015bifurcation}, which justifies the presence of the second term on the right-hand side of Eq. (\ref{M12}). Consequently, the flux of f-cytokines through the endothelium into the intima is influenced by the increasing concentrations of both oxLDL cholesterol and f-cytokines at the endothelium \citep{hansson2006immune}. \\

\noindent
On the moving plaque boundary $\Gamma_2(t)$, we assume continuity of stress so that total stress of inflammatory and non-inflammatory cell phase balances the stress of the outer media region, which is supposed to be constant. Accordingly, we set
\begin{equation}\label{M13}
\left(\mathbb{T}^{f} + \mathbb{T}^{n}\right)|_{\Gamma_2(t)} = \mathbb{T}^{\infty}.
\end{equation}
We also assume that outside the plaque boundary $\Gamma_2(t)$, oxLDL cholesterol, intracellular cholesterol and f-cytokines concentration are kept at constant sources, and thus we set
\begin{equation}\label{M14}
c_1\big|_{\Gamma_2(t)} = c_1^{\infty},
\end{equation}
\begin{equation}\label{M14_2}
 c_2\big|_{\Gamma_2(t)} = c_2^{\infty},
\end{equation}
\begin{equation}\label{M15}
 c_3\big|_{\Gamma_2(t)} = c_3^{\infty}.
\end{equation}
Intracellular cholesterol is formed due to the internalization of extracellular oxLDL upon ingestion. Therefore, at the plaque outer boundary $\Gamma_2(t)$, it can be considered that $c_2^{\infty} = r_{c}c_1^{\infty}$, where $r_c$ is the ratio of internalized cholesterol to extracellular oxLDL cholesterol concentration at that boundary ($0<r_c\leq1$). \\

\noindent
We assume that the boundary $\Gamma_2(t)$ of the plaque expands at the same rate as the normal velocity of the inflammatory cells on that boundary. Therefore, we define the following:
\begin{equation}\label{M16}
\frac{d\Gamma_2}{dt} = \left(\mathbf{v}^{f}.\hat{\mathbf{n}}\right)\big|_{\Gamma_2(t)}.
\end{equation}
Here $\mathbf{v}^{f}$ needs to be determined as part of the solution.
\subsection{\textbf{Constitutive Assumptions}}
\subsubsection{Inflammatory Cell Phase Source Term}
\noindent
Macrophages migrate into the intima in response to f-cytokines. Experimental studies support the notion that f-cytokines, particularly \emph{MCP-1}, stimulate macrophage proliferation during the progression of plaque formation \mbox{\citep{zheng2016local}}. Foam cells are produced when macrophages consume oxLDL present in the non-inflammatory phase. Therefore, the production of inflammatory cells depends on the concentration of oxLDL cholesterol ($c_1$) and f-cytokines ($c_3$). On the other hand, inflammatory cell death occurs due to the toxic effects of an excessive concentration of intracellular cholesterol ($c_2$). In addition, inflammatory cells undergo decay through natural cell death. Therefore, we choose $\daleth^{f}$ by assuming both production and degradation or death saturating rates following Michaelis-Menten kinetics
\begin{equation}\label{M17}
 \daleth^{f} = \underbrace{\varphi^{f} \varphi^{n} \left(\frac{s_0 c_3}{1+s_1 c_3} + \frac{s_2 c_1}{1+s_3 c_1}\right)}_{\text{production}} - \underbrace{\varphi^{f} \left(\frac{s_4 + s_5 c_2}{1+s_6 c_2}\right)\mathcal{H}(c_2-c^{*}_2)}_{\text{death}}-\underbrace{\varphi^{f} s_7}_{\text{decay}},
\end{equation}
where the first two terms correspond to the production of inflammatory cells characterized by the parameters $s_0$, $s_1$, $s_2$, and $s_3$, and the third term represents the death of inflammatory cells characterized by $s_4$, $s_5$, and $s_6$. The production rate of inflammatory cells increases with $c_1$. Upon uptake, extracellular oxLDL is internalized into intracellular cholesterol. Excessive intracellular cholesterol ($c_2$) becomes toxic, leading to cell death. Consequently, a critical threshold concentration of intracellular cholesterol $c^{*}_{2}$ (referred to as the cholesterol-induced reference toxicity level) exists, beyond which inflammatory cells experience toxicity-induced death. Such a choice would be biologically more relevant.  We quantify this phenomenon using the Heaviside function ($\mathcal{H}$).  Beyond the threshold, the death rate of inflammatory cells increases with $c_2$. The last term denotes a marginal linear decay component corresponding to natural death of inflammatory cells. While generating the numerical results, in place of standard form of the Heaviside function $\mathcal{H}$ in Eq. \mbox{(\ref{M17})}, we approximate it by a smooth function as \mbox{\citep{hubbard2013multiphase}}
\begin{equation}
 \mathcal{H}(x,\epsilon) = \frac{1}{2} \left\{1+\tanh\left(\frac{x}{\epsilon}\right)\right\},~~~~~~ \epsilon\ll1 .
\end{equation}
\subsubsection{OxLDL cholesterol, intracellular cholesterol and f-cytokines source terms}
\noindent
Net source or sink of oxLDL cholesterol, $Q_1$, and intracellular cholesterol, $Q_2$, are assumed in the following forms
\begin{subequations}
\begin{equation}\label{M18}
 Q_1 = \underbrace{\varphi^{f} \varphi^n \left(\frac{q_0 c_1}{1+q_1 c_1}\right)}_{\text{consumption}},
\end{equation}
\begin{equation}\label{M18_1}
 Q_2 = - \underbrace{\varphi^{f} \varphi^n \left(\frac{q_0 c_1}{1+q_1 c_1}\right)}_{\text{production}} + \underbrace{\varphi^{f}\varphi^{n} (q_2 c_2)}_{\text{loss in cell death}},
\end{equation}
\end{subequations}
in which $q_0$, $q_1$ and $q_2$ are positive constants. The term on the right-hand side of equation \mbox{\eqref{M18}} describes the loss of oxLDL cholesterol due to its consumption by macrophages to become foam cells. We use a Michaelis-Menten general kinetics type function to represent it. Upon ingestion, extracellular oxLDL cholesterol is directly internalized into intracellular cholesterol. Hence, the term in equation \mbox{\eqref{M18}} accounting for the loss of oxLDL cholesterol can be considered equivalent to the gain or production of intracellular cholesterol, as described in equation \mbox{\eqref{M18_1}} \mbox{\citep{ahmed2024hdl}}. Cell death leads to the release of intracellular cholesterol into the dead materials, a part of the non-inflammatory cell phase. The second term on the right-hand side of equation \mbox{\eqref{M18_1}} accounts for this loss of intracellular cholesterol due to cell death. Note that there may be a loss of oxLDL, perhaps through degradation, but we do not include this for simplicity.\\

\noindent
Similarly, $Q_3$, the net source or sink of f-cytokines can be set as
\begin{equation}\label{M19}
 Q_3 = \underbrace{\varphi^n q_3 c_3}_{\text{decay}} - \underbrace{\varphi^n\varphi^f (q_4 c_1 c_3)}_{\text{production}} ,
\end{equation}
where $q_3$ and $q_4$ are positive constants. Cytokines secreted from foam cells in presence of oxLDL \citep{chalmers2017nonlinear}. Correspondingly, the production of f-cytokines depends on $\varphi^f$ and $c_1$, as shown by the second term of the right hand side of above equation. The first term represents a linear decay in $c_3$.
\subsubsection{Terms Contributing to the Stress Tensor}
\noindent
The function $\Sigma$ in Eq. (\ref{M8a}) represents the pressure on inflammatory cell phase due to cell-cell interactions, which can be expressed as
\begin{equation}\label{M20}
\Sigma\left(\varphi^{f}\right) = \frac{\left(\varphi^{f}-\varphi^{*}\right)}{\left(1-\varphi^{f}\right)^2} \mathcal{H}(\varphi^{f} - \varphi^{*}) ,
\end{equation}
where $\varphi^{*}$ is the volume fraction of foam cells and macrophages beyond which cells experience stress. Our choice of $\Sigma$ follows the study of \citet{breward2002role,byrne2003two}. f-Cytokines act as chemoattractants to pull monocytes from the lumen into the intima. It has already been mentioned that monocytes rapidly transform into macrophages inside the intima. Hence, the movement of macrophages within the intima can be assumed to be influenced by f-cytokines concentration (i.e. chemotaxis). Therefore, $\Lambda$ in Eq. (\ref{M8a}) can defined as
\begin{equation}\label{M21}
 \Lambda = \frac{\chi}{1+(\kappa c_3)^m} ,
\end{equation}
where $\chi$, $\kappa$ and $m$ are positive parameters. The assumption implies that the stress from the inflammatory phase decreases with increasing f-cytokines concentration, allowing the chemoattractant to move. Correspondingly, it is represented as a mechanism of stress relief in the inflammatory cell phase \citep{watson2018two}. The chemotactic function $\Lambda$ also can be defined as $\chi/(1+\kappa c_3)^2$, but in the case of the receptor binding, it may be more reasonable to describe by Hill function as shown in (\ref{M21}) \citep{hillen2009user}. In this model, monocytes enter the intima from the bloodstream with the help of receptors VCAM-1 and ICAM-1. Consequently, our study would prefer to consider the chemotaxis function as (\ref{M21}).
\section{1D Model}
\noindent
As the first step, we study plaque growth in one dimension along the $x$-axis. It occupies the region $0\leq x \leq l(t)$ (see Fig. \ref{athero}). Consequently, the boundary $\Gamma_1$ reduces to $x=0$ (i.e., the interface of the endothelium and intima), and the moving edge $\Gamma_2(t)$ corresponds to the interface of the intima. IEL is now $x=l(t)$, which is the width of the growing plaque at time $t$ in the early stage. This one-dimensional model would enable us to focus on the cellular and biochemical mechanisms without worrying about the geometry of the domain. Furthermore, the diameter of an artery is significantly larger than the width of the intima at the early stages of plaque formation \citep{bonithon1996relation}, which makes it reasonable to use Cartesian coordinates as an approximation for radial coordinates. We denote all the physical variables consistent with the previous section. The above description enables us to reduce the mass balance equations (\ref{M1}) into the following form:
\begin{equation}\label{O1}
\frac{\partial \varphi^{f}}{\partial t} +\frac{\partial}{\partial x} \left(\varphi^{f} v^f \right) = \daleth^{f} ,
\end{equation}
\begin{equation}\label{O2}
\frac{\partial \varphi^{n}}{\partial t} +\frac{\partial}{\partial x} \left(\varphi^{n} v^n \right) = \daleth^{n}.
\end{equation}
Similarly, the momentum conservation equations (\ref{M4}) reduce as,
\begin{equation}\label{O3}
 \frac{\partial}{\partial x} \left(\varphi^{f} T^{f}\right) + \Pi^{f} = 0 ,
\end{equation}
\begin{equation}\label{O4}
 \frac{\partial}{\partial x} \left(\varphi^{n} T^{n}\right) + \Pi^{n} = 0,
\end{equation}
in which
\begin{equation}\label{O5}
\Pi^{f} = p\frac{\partial \varphi^{f}}{\partial x} + k \varphi^{f} \varphi^{n} (v^{n}-v^{f})
\end{equation}
and
\begin{equation}\label{O6}
\Pi^{n} = p\frac{\partial \varphi^{n}}{\partial x} + k \varphi^{f} \varphi^{n} (v^{f}-v^{n}),
\end{equation}
as obtained from (\ref{M7}). The stresses described in Eqs. (\ref{M8a}) and (\ref{M8b}) are reduced to the following 1D form
\begin{equation}\label{O7}
 T^f = - \left(p  + \Sigma\left(\varphi^{f}\right) + \Lambda(c_3)\right)
\end{equation}
and
\begin{equation}\label{O8}
T^n = - p
\end{equation}
respectively. \\

\noindent
Next, the mass transport equations (\ref{M8})-(\ref{M8_1}) reduce to
\begin{equation}\label{O9}
D_1 \frac{\partial}{\partial x}\left(\varphi^n \frac{\partial c_1}{\partial x} \right) = Q_1 ,
\end{equation}
\begin{equation}\label{O9_1}
D_2 \frac{\partial}{\partial x}\left(\varphi^f \frac{\partial c_2}{\partial x} \right) = Q_2 ,
\end{equation}
\begin{equation}\label{O10}
D_3 \frac{\partial}{\partial x}\left(\varphi^n \frac{\partial c_3}{\partial x} \right) = Q_3.
\end{equation}
Finally, The initial and boundary conditions (\ref{M9})-(\ref{M16}) become
\begin{equation}\label{O11}
\varphi^f = \varphi^{f}_0, ~~ l=l_0 , ~~~~~ \text{at}~ t=0,
\end{equation}
\begin{equation}\label{O12}
 \varphi^f v^f = \Xi(c_3), ~~~~~ \text{at}~ x=0,
\end{equation}
\begin{equation}\label{O13}
D_1\left(\frac{\partial c_1}{\partial x}\right) = - \nu_1,  ~~~~~ \text{at}~ x=0,
\end{equation}
\begin{equation}\label{O13_1}
 D_2\left(\frac{\partial c_2}{\partial x}\right) = 0,  ~~~~~ \text{at}~ x=0,
\end{equation}
\begin{equation}\label{O14}
 D_3\left(\frac{\partial c_3}{\partial x}\right) = - \nu_2 c_1 - \nu_3 c_3,  ~~~~~ \text{at}~ x=0,
\end{equation}
\begin{equation}\label{O15}
T^{f}+T^{n} = T^{\text{con}}, ~~~~~ \text{at}~ x=l(t),
\end{equation}
\begin{equation}\label{O16}
c_1 = c^{\infty}_{1}, ~~~~~ \text{at}~ x=l(t),
\end{equation}
\begin{equation}\label{O16_1}
c_2 = r_{c}c^{\infty}_{1}, ~~~~~ \text{at}~ x=l(t),
\end{equation}
\begin{equation}\label{O17}
 c_3 = c^{\infty}_{3}, ~~~~~ \text{at}~ x=l(t),
\end{equation}
and
\begin{equation}\label{O18}
\frac{dl}{dt} = v^f. ~~~~~ \text{at}~ x=l(t).
\end{equation}
\subsection{\textbf{Model Simplification}}
\noindent
The entire model of early stage plaque growth is expressed as a system of nonlinear partial differential equations in $\varphi^f$, $v^f$, $c_1$, $c_2$ and $c_3$. The simplification of this model follows the similar approach to that of \citet{breward2002role,byrne2003modelling,watson2018two}, and for clarity, the model simplification is included in detail. First we add the mass balance equations (\ref{O1}) and (\ref{O2}), to get
\begin{equation}\label{S1}
\frac{\partial}{\partial x}\left(\varphi^f v^f+\varphi^n v^n\right)= 0,
\end{equation}
which upon integration and imposing $v^{\text{com}}=\varphi^f v^f+\varphi^n v^n = 0$ at $x=0$, we obtain
\begin{equation}\label{S2}
v^n = - \left(\frac{\varphi^f}{\varphi^n}\right) v^f.
\end{equation}
The vanishing composite velocity of the mixture at the endothelium-intima interface indicates that, through this interface, the influx of inflammatory tissue equates to the outflux of non-inflammatory tissue. A detailed discussion of this context is provided later. \\

\noindent
Now, summing up the momentum balance equations (\ref{O3}) and (\ref{O4}), and upon submitting the stress expressions $T^f$ and $T^n$ from (\ref{O7})-(\ref{O8}) respectively into the resultant, we obtain
\begin{equation}\label{S3}
\frac{\partial p}{\partial x} + \frac{\partial F}{\partial x} = 0,
\end{equation}
where
\begin{equation}\label{S4}
F=\varphi^f(\Sigma+\Lambda).
\end{equation}
We derive the following from (\ref{S3}) with the help of (\ref{O3}), (\ref{O5}) and (\ref{O7}):
\begin{equation}\label{S5}
\varphi^n \frac{\partial F}{\partial x} + k \varphi^f v^f = 0.
\end{equation}
Consequently, with the help of mass balance equation (\ref{O1}), we deduce
\begin{equation}\label{S6}
\frac{\partial \varphi^f}{\partial t} - \frac{\varphi^n}{k}\left[\frac{\partial^2 F}{\partial x^2}+\frac{1}{\varphi^n}\frac{\partial \varphi^n}{\partial x}\frac{\partial F}{\partial x} \right]= \daleth^{f}.
\end{equation}
\noindent
Now $F$ from Eq. (\ref{S4}) is substituted into Eq. (\ref{S6}) to derive
\begin{equation}\label{SN1}
\frac{\partial \varphi^f}{\partial t} = \frac{\partial}{\partial x} \left(\frac{\varphi^n}{k}\frac{\partial}{\partial x} \left(\varphi^f \Sigma + \varphi^f \Lambda \right)\right) + \daleth^{f},
\end{equation}
which is further manipulated to
\begin{equation}\label{SN2}
 \frac{\partial \varphi^f}{\partial t} = \frac{\partial}{\partial x}\left(D^f \frac{\partial \varphi^f}{\partial x} - C^f \varphi^f \frac{\partial c_3}{\partial x} \right) + \daleth^{f},
\end{equation}
where
\begin{equation}\label{SN3}
 D^f = \frac{\varphi^n}{k} \left(\Sigma +  \varphi^f \frac{\partial \Sigma}{\partial \varphi^f} +  \Lambda \right)
\end{equation}
 is the effective nonlinear diffusivity, and
\begin{equation}\label{SN4}
 C^f = - \frac{\varphi^n}{k} \frac{d\Lambda}{dc_3} = \frac{\varphi^n}{k} \left( \frac{\chi m \kappa^m {c_3}^{m-1}}{(1+(\kappa c_3)^m)^2}\right)
\end{equation}
is the effective nonlinear chemotaxis coefficient. Note that the chemotaxis coefficient depends on the f-cytokines concentration $c_3$. Therefore, the overall chemotactic flux at any point in the domain relies on both the local gradient and the specific concentration of f-cytokines.\\

\noindent
As mentioned in the boundary condition (\ref{M10}), inflammatory cells are assumed to enter the intima via a chemotactic response to the f-cytokines at the endothelium-intima interface boundary, so that,
\begin{equation}\label{SN5}
 \varphi^f v^f = C^f \varphi^f_{\#} \frac{\partial c_3}{\partial x} ~~~~~~ \text{at}~~~ x=0,
\end{equation}
which in equivalent form is given by
\begin{equation}\label{SN6}
 \frac{\partial F}{\partial x} = \frac{d\Lambda}{dc_3} \varphi^f_{\#} \frac{\partial c_3}{\partial x} ~~~~~~ \text{at}~~~ x=0,
\end{equation}
where $\varphi^f_{\#}$ is a constant volume fraction of monocytes within the lumen which are ready to enter the intima. Consequently, the expression for $\Xi(c_3)$ takes the following form:
\begin{equation}\label{SN7}
 \Xi(c_3) = - \frac{\varphi^n}{k} \frac{d\Lambda}{dc_3} \varphi^f_{\#} \frac{\partial c_3}{\partial x} = \frac{\varphi^n}{k} \left( \frac{\chi m \kappa^m {c_3}^{m-1}}{(1+(\kappa c_3)^m)^2}\right) \varphi^f_{\#} \frac{\partial c_3}{\partial x},
\end{equation}
Thus, the boundary condition (\ref{O12}) reduces to the condition (\ref{SN6}), and upon substituting $F$ into it, we obtain
\begin{equation}\label{SN8}
 \frac{\partial }{\partial x} \left(\varphi^f \Sigma\right) + \frac{\partial}{\partial x}\left(\varphi^f \Lambda\right) = \frac{d\Lambda}{dc_3} \varphi^f_{\#} \frac{\partial c_3}{\partial x}  ~~~~~~ \text{at}~~~ x=0.
\end{equation}
Now it is reasonable to assume that at the interface of the endothelium and intima (i.e. at $x=0$), inflammatory cells do not interact with each other so that
\begin{equation}\label{SN9}
 \frac{\partial}{\partial x}\left(\varphi^f \Lambda\right) = \frac{d\Lambda}{dc_3} \varphi^f_{\#} \frac{\partial c_3}{\partial x} ~~~~~~ \text{at}~~~ x=0.
\end{equation}
Note that the structure of this boundary chemotactic flux is consistent with the terms derived in equations (\ref{SN2}) and (\ref{SN4}). The model assumes that inflammatory cells enter the intima in response to f-cytokines. In other words, the lack of f-cytokines at the endothelium-intima interface would prevent inflammatory cells from entering the intima, as stipulated by the condition. In addition, this boundary condition indicates that an equivalent efflux from the non-inflammatory phase must balance the influx of inflammatory cells into our closed system. Physically, this efflux is a transfer of interstitial fluid towards the lumen. Although studies suggest that there may be an efflux of water from the intima towards the lumen under certain conditions \citep{joshi2020pre}, the assumption that a non-inflammatory phase efflux will exactly balance the inflammatory cell influx is an oversimplification. However, the model remains reasonable, provided the total influx of monocytes remains relatively small. Indeed, an \emph{in-vitro} experimental study on human saphenous veins indicates that the proportion of monocytes adhering to the endothelium is around 10\% \citep{cooper1991monocyte}, thus supporting the notion of marginal influx through the endothelium. \\

\noindent
The stress boundary condition (\ref{O15}) reduces to (after scaling with $T^{\text{con}}$, which is later mentioned in the nondimensionalization section),
\begin{equation}\label{S22}
T^{f}+T^{n}=1,  ~~~~~ \text{at}~ x=l(t),
\end{equation}
which further simplifies to at $x=l(t)$:
\begin{equation}\label{S23}
\Sigma(\varphi^f)+\Lambda(c_3)+(2p+1)=0.
\end{equation}
At the plaque boundary $x=l(t)$, we assume the interstitial hydrodynamic pressure in such a way that $2p+1=0$. We therefore have
\begin{equation}\label{S24}
\Sigma(\varphi^f)+\Lambda(c_3)=0, ~~~~~ \text{at}~~ x=l(t).
\end{equation}
As a result, our reduced model can be read as the Eq. \mbox{(\ref{O9})} coupled with Eqs. \mbox{(\ref{O9_1})}, \mbox{(\ref{O10})}, \mbox{(\ref{S5})} and \mbox{(\ref{S6})} subject to the initial and boundary conditions prescribed by \mbox{(\ref{O11})}, \mbox{(\ref{O13})}-\mbox{(\ref{O14})}, \mbox{(\ref{O16})}-\mbox{(\ref{O18})}, \mbox{(\ref{SN9})} and \mbox{(\ref{S24})}.
\subsection{\textbf{Nondimensionalization}}
\noindent
Using hat (`~$\hat{}$~') to denote dimensionless quantities, we scale the independent and dependent variables $x$, $t$, $l$, $v^f$, $c_1$, $c_2$, $c_3$, $\varphi^f$ and $\varphi^n$, as follows (one can note that, inflammatory and non-inflammatory cell phase volume fractions $\varphi^f$ and $\varphi^n$ respectively do not require scaling):
\begin{equation*}
\hat{x}=x/l_0 ~,~~~ \hat{t}=t/\left(kl^2_{0}/T^{\infty}\right) ~,~~~ \hat{l}=l/l_0 ~,~~~ \hat{v}^{f}=v^f/\left(T^{\infty}/kl_0 \right).
\end{equation*}
\begin{equation*}
\hat{c_1}=c_1/c^{\infty}_{1} ~,~~ \hat{c}_{2}=c_2/c^{\infty}_{1} ~,~~ \hat{c}_{3}=c_3/c^{\infty}_{3} ~,~~ \hat{\varphi}^f = \varphi^f ~,~~ \hat{\varphi}^n = \varphi^n.
\end{equation*}
We introduced the non-dimensional parameters in the following way:
\noindent
\begin{equation*}
\hat{\daleth^{f}} = \daleth^{f}/\left(T^{\infty}/kl^{2}_0 \right) ~,~~ \hat{\Sigma} = \Sigma/ T^{\infty} ~,~~ \hat{\Lambda} = \Lambda/ T^{\infty}~,~~\hat{T}^{f} = T^f/ T^{\infty}~,~~ \hat{T}^{n} = T^{n}/ T^{\infty},
\end{equation*}
\begin{equation*}
\hat{F} = F/T^{\infty} ~,~~ \hat{Q}_1=Q_1/\left(D_1 c^{\infty}_1/ l^{2}_0\right)~,~~ \hat{Q}_{2}=Q_2/\left(D_2 c^{\infty}_1/l^{2}_0 \right)~,~~ \hat{Q}_{3}=Q_3/\left(D_3 c^{\infty}_3/l^{2}_0 \right),
\end{equation*}
\begin{equation*}
\hat{q_{0}} = q_{0}/\left(D_{1}/l_{0}^{2}\right) ~,~~~ \hat{q_{1}} = q_{1}/\ \left(1/c_{1}^{\infty}\right) ~,~~~ \hat{q_{2}} = q_{2}/\left(D_{2}/l_{0}^{2}\right) ~ ,~~~ \hat{q_{3}} = q_{3}/\left(D_{3}/l_{0}^{2}\right) ~,
\end{equation*}
\begin{equation*}
 \hat{q_{4}} = q_{4}/\left(D_{3}/l_{0}^{2} c_{1}^{\infty}\right) ~,~~~ \hat{s_{0}} = s_{0}/\left(T^{\infty}/kl_{0}^{2}c_{3}^{\infty}\right) ~,~~~ \hat{s_{1}} = s_{1}/\left(1/c_{3}^{\infty}\right) ~,
\end{equation*}
\begin{equation*}
 \hat{s_{2}} = s_{2}/\left(T^{\infty}/kl_{0}^{2}c_{1}^{\infty}\right) ~,~~~ \hat{s_{3}} = s_{3}/\left(1/c_{1}^{\infty}\right) ~,~~~ \hat{s_{4}} = s_{4}/\left(T^{\infty}/kl_{0}^{2}\right) ~,
\end{equation*}
\begin{equation*}
   \hat{s_{5}} = s_{5}/\left(T^{\infty}/kl_{0}^{2}c_{1}^{\infty}\right) ~,~~~ \hat{s_{6}} = s_{6}/\left(1/c_{1}^{\infty}\right) ~,~~~ \hat{s_{7}} = s_{7}/\left(T^{\infty}/kl_{0}^{2}\right)~,
\end{equation*}
\begin{equation*}
\hat{\nu}_1=\nu_1/\left(D_1 c^{\infty}_1/l_0\right) ~,~~~~ \hat{\nu}_{2}=\nu_2/\left(D_3 c^{\infty}_3/l_0 c^{\infty}_1\right) ~,~~~~ \hat{\nu}_3=\nu_3/(1/l_0).
\end{equation*}
\noindent
The corresponding model equations, initial and boundary conditions in non-dimensional form (removing hat notation (`~$\hat{}$~') for convenience and without loss of generality) reduce to \\
\begin{equation}\label{S7}
\frac{\partial \varphi^{f}}{\partial t} - \varphi^{n}\frac{\partial^2 F}{\partial x^2} - \frac{\partial \varphi^{n}}{\partial x}\frac{\partial F}{\partial x} = \daleth^{f}
\end{equation}
\begin{equation}\label{S8}
\varphi^{n} \frac{\partial F}{\partial x} + \varphi^{f} v^{f} = 0,
\end{equation}
\begin{equation}\label{S9}
\frac{\partial}{\partial x}\left(\varphi^{n} \frac{\partial c_{1}}{\partial x} \right) = Q_{1} ,
\end{equation}
\begin{equation}\label{S9_1}
 \frac{\partial}{\partial x}\left(\varphi^{f} \frac{\partial c_{2}}{\partial x} \right) = Q_{2},
\end{equation}
\begin{equation}\label{S10}
 \frac{\partial}{\partial x}\left(\varphi^{n} \frac{\partial c_{3}}{\partial x} \right) = Q_{3},
\end{equation}
\begin{equation}\label{S11}
 \varphi^{f} = \varphi^{f}_0, ~~ l=1 , ~~~~~ \text{at}~~ t=0,
 \end{equation}
\begin{equation}\label{S12}
 \frac{\partial}{\partial x}\left(\varphi^{f} \Lambda \right) = \frac{d\Lambda}{dc_3} \varphi^f_{\#} \frac{\partial c_3}{\partial x} ~~~~~~ \text{at}~~~ x=0,
\end{equation}
\begin{equation}\label{S13}
\frac{\partial c_{1}}{\partial x} = - \nu_{1},  ~~~~~ \text{at}~ x=0,
\end{equation}
\begin{equation}\label{S13_1}
 \frac{\partial c_{2}}{\partial x} = 0,  ~~~~~ \text{at}~ x=0,
\end{equation}
\begin{equation}\label{S14}
 \frac{\partial c_{3}}{\partial x} = - \nu_{2} c_{1} - \nu_{3} c_{3},  ~~~~~ \text{at}~ x=0,
\end{equation}
\begin{equation}\label{S15}
\Sigma(\varphi^{f})+\Lambda(c_{3})=0, ~~~~~ \text{at}~~ x=l(t),
\end{equation}
\begin{equation}\label{S16}
c_{1} = 1, ~~~~~ \text{at}~ x=l(t),
\end{equation}
\begin{equation}\label{S16_1}
 c_{2} = r_c, ~~~~~ \text{at}~ x=l(t),
\end{equation}
\begin{equation}\label{S17}
 c_3 = 1, ~~~~~ \text{at}~ x=l(t),
\end{equation}
and
\begin{equation}\label{S18}
\frac{dl}{dt} = v^{f}. ~~~~~ \text{at}~ x=l(t).
\end{equation}
\subsection{\textbf{Change of Variables}}
\noindent
In order to execute numerical solutions, it is convenient to map the moving domain onto a fixed one. Consequently, we change the variables $(x,t)$ to $(\zeta, \tau)$ by the transformation $\zeta=x/l(t)$ and $\tau=t$. The transformed problem reduces to
\begin{equation}\label{S25}
\frac{\partial \varphi^f}{\partial \tau} - \frac{\zeta}{l}\frac{dl}{d\tau} \frac{\partial \varphi^f}{\partial \zeta}- \frac{\varphi^n}{l^2} \frac{\partial^2 F}{\partial \zeta^2} - \frac{1}{l^2} \frac{\partial \varphi^n}{ \partial \zeta} \frac{\partial F}{ \partial \zeta} = \daleth^{f},
\end{equation}
\begin{equation}\label{S26}
\varphi^n \frac{\partial F}{\partial \zeta} + l\varphi^f v^f =0,
\end{equation}
\begin{equation}\label{S27}
\frac{\partial}{\partial \zeta}\left(\varphi^n \frac{\partial c_1}{\partial \zeta} \right) = l^2 Q_1,
\end{equation}
\begin{equation}\label{S27_1}
 \frac{\partial}{\partial \zeta}\left(\varphi^f \frac{\partial c_2}{\partial \zeta} \right) = l^2 Q_2,
\end{equation}
\begin{equation}\label{S28}
 \frac{\partial}{\partial \zeta}\left(\varphi^n \frac{\partial c_3}{\partial \zeta} \right) = l^2 Q_3,
\end{equation}
with initial and boundary conditions
\begin{equation}\label{S29}
 \varphi^f = \varphi^f_0, ~~ l=1 , ~~~~~ \text{at}~~ \tau=0,
\end{equation}
\begin{equation}\label{S30}
 \frac{\partial}{\partial \zeta} \left(\varphi^f \Lambda \right) = \frac{d\Lambda}{dc_3} \varphi^f_{\#} \frac{\partial c_3}{\partial \zeta},  ~~~~ \text{at}~~ \zeta=0,
\end{equation}
\begin{equation}\label{S31}
\frac{\partial c_1}{\partial \zeta} = - l \nu_1,  ~~~~~ \text{at}~ \zeta=0,
\end{equation}
\begin{equation}\label{S31_1}
 \frac{\partial c_2}{\partial \zeta} = 0,  ~~~~~ \text{at}~ \zeta=0,
\end{equation}
\begin{equation}\label{S32}
 \frac{\partial c_3}{\partial \zeta} = - l \nu_2 c_1 - l \nu_3 c_3,  ~~~~~ \text{at}~ \zeta=0,
\end{equation}
\begin{equation}\label{S33}
\Sigma(\varphi^f)+\Lambda(c_3)=0, ~~~~~ \text{at}~~ \zeta=1,
\end{equation}
\begin{equation}\label{S34}
c_1 = 1, ~~~~~ \text{at}~~ \zeta=1,
\end{equation}
\begin{equation}\label{S34_1}
 c_2 = r_c, ~~~~~ \text{at}~~ \zeta=1,
\end{equation}
\begin{equation}\label{S35}
 c_3 = 1, ~~~~~ \text{at}~~ \zeta=1,
\end{equation}
and
\begin{equation}\label{S36}
\frac{dl}{d\tau} = v^f ~~~~~ \text{at}~~ \zeta=1.
\end{equation}
\section{Computational Technique}
\noindent
The present differential nonlinear system is parabolic and amenable to using explicit finite‐difference formulation. Accordingly, one can use an explicit finite-difference scheme to develop a numerical solution for the nonlinear system. However, applying such a central difference formula to equation \mbox{\eqref{S27}} leads to a system of nonlinear equations for the unknown $c_1$ at discretized grid points. This nonlinear system of equations requires due attention. In particular, Newton's method can solve such nonlinear system, however, this is a tedious job demanding high level of computational resources and skills. One approach to circumvent this difficulty is to assume $q_1=0$  and remove the nonlinearity \mbox{\citep{breward2002role,lee2013moving}}. But, this restricts the general kinetics and reduces to first-order kinetics, a particular case of Michaelis–Menten kinetics. Hence, in the present study, rather than taking a vanishing limit of $q_1 \rightarrow 0$, we wish to propose a perturbation solution. For this, we deal with the general kinetics and consider $q_1$ as a small parameter ($q_1\ll1$) so that it becomes a perturbation parameter. Therefore, we reduce the nonlinear form of $Q_1$ that depends on $c_1$ to a weakly linearized form using perturbation approximation with the perturbed parameter $q_1$ as described in the following section.
\subsection{\textbf{Perturbation Approximations}}
\noindent
We consider the following perturbation approximations \\
\begin{equation}\label{P1}
c_\ell = c_\ell^{(0)} + q_1 c_\ell^{(1)} + \mathcal{O}(q_1^2), ~~~~~~ \ell = 1,2,3
\end{equation}
\begin{equation}\label{P2}
\varphi^f = {\left(\varphi^{f}\right)}^{(0)} + q_1 {\left(\varphi^{f}\right)}^{(1)} + \mathcal{O}(q^{2}_1), ~~~~~~~~~~~~~~
\end{equation}
\begin{equation}\label{P3}
v^f = {\left(v^{f}\right)}^{(0)} + q_1{\left(v^{f}\right)}^{(1)} + \mathcal{O}(q_1^2). ~~~~~~~~~~~~~~
\end{equation}
Using the above approximations, we can get the following expansions
\begin{equation}\label{P4}
\varphi^{n} = {\left(\varphi^{n}\right)}^{(0)} + q_1{\left(\varphi^{n}\right)}^{(1)} + \mathcal{O}(q_1^2), ~~~~~~~~~~~~~~
\end{equation}
\begin{equation}\label{P5}
\daleth^{f} = {\daleth^{f}}^{(0)} + q_1 {\daleth^{f}}^{(1)} + \mathcal{O}(q_1^2), ~~~~~~~~~~~~~~
\end{equation}
\begin{equation}\label{P6}
\Sigma = \Sigma^{(0)} +q_1 \Sigma^{(1)} +  \mathcal{O}(q_1^2), ~~~~~~~~~~~~~~
\end{equation}
\begin{equation}\label{P7}
\Lambda = \Lambda^{(0)} + q_1 \Lambda^{(1)} + \mathcal{O}(q_1^2), ~~~~~~~~~~~~~~
\end{equation}
\begin{equation}\label{P8}
F = F^{(0)} + q_1 F^{(1)} + \mathcal{O}(q_1^2), ~~~~~~~~~~~~~~
\end{equation}
and
\begin{equation}\label{P9}
Q_\ell = Q_\ell^{(0)} + q_1 Q_\ell^{(1)} + \mathcal{O}(q_1^2), ~~~~~~ \ell = 1,2,3
\end{equation}
where the expressions of ${\left(\varphi^n\right)}^{(0)}$, ${\left(\varphi^n\right)}^{(1)}$, ${\daleth^{f}}^{(0)}$, ${\daleth^{f}}^{(1)}$, $\Sigma^{(0)}$, $\Sigma^{(1)}$, $\Lambda^{(0)}$, $\Lambda^{(1)}$, $F^{(0)}$, $F^{(1)}$, $Q^{(0)}_1$, $Q^{(1)}_1$, $Q^{(0)}_2$, $Q^{(1)}_2$, $Q^{(0)}_3$, and $Q^{(1)}_3$ are described in the Appendix A and B. Correspondingly, the leading and $\mathcal{O}(q_1)$ problems reduce as shown in the following sections.
\subsubsection{The Leading Order Problem}
\noindent
The governing equations and the initial and boundary conditions corresponding to the leading order problem reduce as shown in Table \ref{tableP1}.
\begin{sidewaystable}[htpb]
\caption{Governing equations, along with initial and boundary conditions related to the leading order problem.}
\centering
\begin{tabular}{c}%
\hline
\underline{\textbf{Governing Equations}} \vspace{0.2cm} \\
$\frac{\partial {\left(\varphi^{f}\right)}^{(0)}}{\partial \tau} - \frac{\zeta}{l}\frac{dl}{d\tau}\frac{\partial {\left(\varphi^{f}\right)}^{(0)}}{\partial \zeta} - \frac{{\left(\varphi^{n}\right)}^{(0)}}{l^2} \frac{\partial^2 F^{(0)}}{\partial \zeta^2} - \frac{1}{l^2} \frac{\partial {\left(\varphi^{n}\right)}^{(0)}}{\partial \zeta} \frac{\partial F^{(0)}}{ \partial \zeta} = {\daleth^{f}}^{(0)} ,~~~~~~~~~~\text{(T1.1)} $ \\\\
 $ {\left(\varphi^{n}\right)}^{(0)} \frac{\partial F^{(0)}}{\partial \zeta} + l {\left(\varphi^{f}\right)}^{(0)} {\left(v^{f}\right)}^{(0)} = 0, ~~~~~~~~~~ \text{(T1.2)} $ \\\\
$ \frac{\partial}{\partial \zeta}\left({\left(\varphi^{n}\right)}^{(0)} \frac{\partial c^{(0)}_1}{\partial \zeta} \right) = l^2 Q^{(0)}_1, ~~~~~~~~~~\text{(T1.3)} $ \\\\
$ \frac{\partial}{\partial \zeta}\left({\left(\varphi^{f}\right)}^{(0)} \frac{\partial c^{(0)}_2}{\partial \zeta} \right) = l^2 Q^{(0)}_2, ~~~~~~~~~~\text{(T1.4)} $ \\\\
$ \frac{\partial}{\partial \zeta}\left({\left(\varphi^{n}\right)}^{(0)} \frac{\partial c^{(0)}_3}{\partial \zeta} \right) = l^2 Q^{(0)}_3, ~~~~~~~~~~\text{(T1.5)} $ \\\\
\hline
\underline{\textbf{Initial Conditions}} \vspace{0.1cm} \\
${\left(\varphi^{f}\right)}^{(0)} = \varphi^f_0, ~~ l=1 , ~~~~~ \textrm{at}~~ \tau=0, ~~~~~~~~~~\text{(T1.6)} $ \\\\
\hline
\underline{\textbf{Boundary Conditions}} \vspace{0.1cm} \\
$\frac{\partial}{\partial \zeta}\left[{\left(\varphi^{f}\right)}^{(0)} \Lambda^{(0)}\right] =  \frac{d\Lambda^{(0)}}{dc^{(0)}_3} \varphi^f_{\#} \frac{\partial c^{(0)}_3}{\partial \zeta}, ~~~~ \text{at}~~ \zeta=0, ~~~~~~~~~~\text{(T1.7)} $ \\\\
$\frac{\partial c_1^{(0)}}{\partial \zeta} = - l \nu_1,  ~~~~~ \textrm{at}~ \zeta=0,~~~~~~~~~~\text{(T1.8)} $ \\\\
$\frac{\partial c_2^{(0)}}{\partial \zeta} = 0,  ~~~~~ \textrm{at}~ \zeta=0, ~~~~~~~~~~\text{(T1.9)} $ \\\\
$\frac{\partial c^{(0)}_3}{\partial \zeta} = - l \nu_2 c_1^{(0)} - l \nu_3 c_3^{(0)},  ~~~~~ \textrm{at}~ \zeta=0,~~~~~~~~~~\text{(T1.10)} $ \\\\
$\Sigma^{(0)}\left({\left(\varphi^{f}\right)}^{(0)}\right) + \Lambda^{(0)}\left(c^{(0)}_3\right)=0, ~~~~~ \textrm{at}~~ \zeta=1, ~~~~~~~~~~\text{(T1.11)} $ \\
$c^{(0)}_1 = 1, ~~~~~ \text{at}~~ \zeta=1, ~~~~~~~~~~\text{(T1.12)} $ \\
$c^{(0)}_2 = r_c, ~~~~~ \text{at}~~ \zeta=1, ~~~~~~~~~~\text{(T1.13)} $ \\
$c^{(0)}_3 = 1, ~~~~~ \text{at}~~ \zeta=1,~~~~~~~~~~\text{(T1.14)} $ \\
$\frac{dl}{d\tau} = {\left(v^{f}\right)}^{(0)} ~~~~~ \textrm{at}~~ \zeta=1. ~~~~~~~~~~\text{(T1.15)} $ \\\\
\hline
\end{tabular}
\label{tableP1}
\end{sidewaystable}
\begin{sidewaystable}[htpb]
\caption{Governing equations, along with initial and boundary conditions corresponding to the $\mathcal{O}(q_1)$ problem.}
\centering
\begin{tabular}{c}%
\hline
\underline{\textbf{Governing Equations}} \vspace{0.2cm} \\
$\frac{\partial {\left(\varphi^{f}\right)}^{(1)}}{\partial \tau} - \frac{\zeta}{l}\frac{dl}{d\tau}\frac{\partial {\left(\varphi^{f}\right)}^{(1)}}{\partial \zeta} -\frac{{\left(\varphi^{n}\right)}^{(0)}}{l^2} \frac{\partial^2 F^{(1)}}{\partial\zeta^2} - \frac{{\left(\varphi^{n}\right)}^{(1)}}{l^2} \frac{\partial^2 F^{(0)}}{\partial\zeta^2} - \frac{1}{l^2} \frac{\partial {\left(\varphi^{n}\right)}^{(0)}}{\partial \zeta} \frac{\partial F^{(1)}}{\partial \zeta} - \frac{1}{l^2} \frac{\partial {\left(\varphi^{n}\right)}^{(1)}}{\partial \zeta} \frac{\partial F^{(0)}}{\partial \zeta} = {\daleth^{f}}^{(1)}, ~~~~~~\text{(T2.1)} $ \\\\
${\left(\varphi^{n}\right)}^{(0)} \frac{\partial F^{(1)}}{\partial \zeta} + {\left(\varphi^{n}\right)}^{(1)} \frac{\partial F^{(0)}}{\partial \zeta} + l {\left(\varphi^{f}\right)}^{(0)} {\left(v^{f}\right)}^{(1)} + l{\left(\varphi^{f}\right)}^{(1)}{\left(v^{f}\right)}^{(0)} = 0, ~~~~~~\text{(T2.2)} $ \\\\
$\frac{\partial}{\partial \zeta}\left({\left(\varphi^{n}\right)}^{(0)} \frac{\partial c_1^{(1)}}{\partial \zeta} + {\left(\varphi^{n}\right)}^{(1)} \frac{\partial c_1^{(0)}}{\partial \zeta} \right) = l^2 Q_1^{(1)}, ~~~~~~\text{(T2.3)} $ \\\\
$\frac{\partial}{\partial \zeta}\left({\left(\varphi^{f}\right)}^{(0)} \frac{\partial c_2^{(1)}}{\partial \zeta} + {\left(\varphi^{f}\right)}^{(1)} \frac{\partial c^{(0)}_2}{\partial\zeta} \right) = l^2 Q_2^{(1)}, ~~~~~~\text{(T2.4)} $ \\\\
$\frac{\partial}{\partial \zeta}\left({\left(\varphi^{n}\right)}^{(0)} \frac{\partial c_3^{(1)}}{\partial \zeta} + {\left(\varphi^{n}\right)}^{(1)} \frac{\partial c^{(0)}_3}{\partial\zeta} \right) = l^2 Q_3^{(1)}, ~~~~~~\text{(T2.5)} $ \\\\
\hline
\underline{\textbf{Initial Conditions}} \vspace{0.1cm} \\
 $ {\left(\varphi^{f}\right)}^{(1)} = 0, ~~ l=1 , ~~~~~ \text{at}~~ \tau=0, ~~~~~~\text{(T2.6)} $ \\\\
\hline
\underline{\textbf{Boundary Conditions}} \vspace{0.1cm} \\
$\frac{\partial}{\partial \zeta} \left({\left(\varphi^{f}\right)}^{(0)} \Lambda^{(1)} + {\left(\varphi^{f}\right)}^{(1)} \Lambda^{(0)} \right) = \frac{d\Lambda^{(0)}}{dc^{(0)}_3} \varphi^f_{\#} \frac{\partial c^{(1)}_3}{\partial \zeta} + \frac{d\Lambda^{(1)}}{dc^{(0)}_3} \varphi^f_{\#} \frac{\partial c^{(0)}_3}{\partial \zeta}, ~~~ \text{at}~~ \zeta=0, ~~~~~~\text{(T2.7)} $  \\\\
$ \frac{\partial c_1^{(1)}}{\partial \zeta} = 0,  ~~~~~ \text{at}~ \zeta=0, ~~~~~~\text{(T2.8)} $ \\\\
$ \frac{\partial c_2^{(1)}}{\partial \zeta} = 0,  ~~~~~ \text{at}~ \zeta=0, ~~~~~~\text{(T2.9)} $ \\\\
$ \frac{\partial c_3^{(1)}}{\partial \zeta} = - l \nu_2 c_1^{(1)} - l \nu_3 c_3^{(1)},  ~~~~~ \text{at}~ \zeta=0, ~~~~~~\text{(T2.10)} $ \\\\
$\Sigma^{(1)} + \Lambda^{(1)} = 0, ~~~~~ \text{at}~~ \zeta=1, ~~~~~~\text{(T2.11)} $ \\\\
$ c^{(1)}_1 = 0, ~~~~~ \text{at}~~ \zeta=1, ~~~~~~\text{(T2.12)} $ \\
$ c^{(1)}_{2} = 0. ~~~~~ \text{at}~~ \zeta=1, ~~~~~~\text{(T2.13)} $ \\
$ c^{(1)}_{3} = 0. ~~~~~ \text{at}~~ \zeta=1. ~~~~~~\text{(T2.14)} $ \\\\
\hline
\end{tabular}
\label{tableP2}
\end{sidewaystable}
\subsubsection{The $\mathcal{O}(q_1)$ Problem}
\noindent
In the first order, the reduced governing equations and the initial and boundary conditions are shown in Table \ref{tableP2}. \\

\noindent
Now, the terms $Q^{(0)}_1$ and $Q^{(1)}_1$ are both linear in $c^{(0)}_1$ and $c^{(1)}_1$ respectively (see Appendix A \& B). Therefore, we can compute $c^{(0)}_{1}$ and $c^{(1)}_1$ numerically from the equations in Table \mbox{\ref{tableP1}} and \mbox{\ref{tableP2}} (specifically, the third governing equations from both tables). Accordingly, we use a stable explicit finite-difference scheme known as the forward-time central space (FTCS) method (refer \mbox{\citet{zaman2016unsteady,lee2013moving}}, \mbox{\citet{hoffmann2000computational}}), to solve the leading order and $\mathcal{O}(q_1)$ Problem. This scheme approximates spatial derivatives by central difference formulae, while the forward difference formula approximates the time derivative.
We discrete the region (0,1) into $N$ equal cells of size $\triangle\zeta = 1/N+1$ with a time step $\triangle\tau>0$. Define $\zeta_j = (j-1)\triangle\zeta$, $j=1,2,...,N+1$, and $\tau_n = (n-1)\triangle\tau$, $n=1,2,...,$ and approximations ${\left[\varphi^{f}\right]}_{j}^n = \varphi^{f}(\zeta_j, \tau_n)$, ${\left[c_1\right]}_{j}^{n} = c_1(\zeta_j, \tau_n)$, ${\left[c_2\right]}_{j}^{n} = c_2(\zeta_j, \tau_n)$ and ${\left[c_3\right]}_{j}^{n} = c_3(\zeta_j, \tau_n)$. We use the following approximations for the spatial and temporal derivatives of volume fractions and concentrations, as given below
\begin{equation}\label{N1}
	\frac{\partial c_\ell}{\partial \zeta} \cong \frac{{\left[c_\ell \right]}^n_{j+1} - {\left[c_\ell \right]}^n_{j-1}}{2\triangle \zeta},~~~~~~~~ \ell=1,2,3
\end{equation}
\begin{equation}\label{N2}
	\frac{\partial^2 c_\ell}{\partial \zeta^2} \cong \frac{{\left[c_\ell\right]}^n_{j+1} -  2{\left[c_\ell\right]}^n_{j}  + {\left[c_\ell\right]}^n_{j-1}}{{\left(\triangle \zeta\right)}^2},~~~~~~~ \ell=1,2,3
\end{equation}
\begin{equation}\label{N3}
	\frac{\partial \varphi^{f}}{\partial \zeta} \cong \frac{{\left[\varphi^{f}\right]}^n_{j+1} - {\left[\varphi^{f}\right]}^n_{j-1}}{2\triangle \zeta},
\end{equation}
\begin{equation}\label{N4}
	\frac{\partial^2 \varphi^{f}}{\partial \zeta^2} \cong \frac{{\left[\varphi^{f}\right]}^n_{j+1} -  2{\left[\varphi^{f}\right]}^n_{j}  + {\left[\varphi^{f}\right]}^n_{j-1}}{{\left(\triangle \zeta\right)}^2},
\end{equation}
and
\begin{equation}\label{N5}
	\frac{\partial \varphi^{f}}{\partial \tau} \cong \frac{{\left[\varphi^{f}\right]}^{n+1}_{j} - {\left[\varphi^{f}\right]}^n_{j}}{\triangle \tau}.
\end{equation}
The above approximations for each variables are generic and these discretizations will be used for the respective variables involved while solving leading order and $\mathcal{O}(q_1)$ problems as shown in Tables \mbox{\ref{tableP1}} and \mbox{\ref{tableP2}}.\\

\begin{figure}[h!]
\centering
\includegraphics[width=1.1\textwidth]{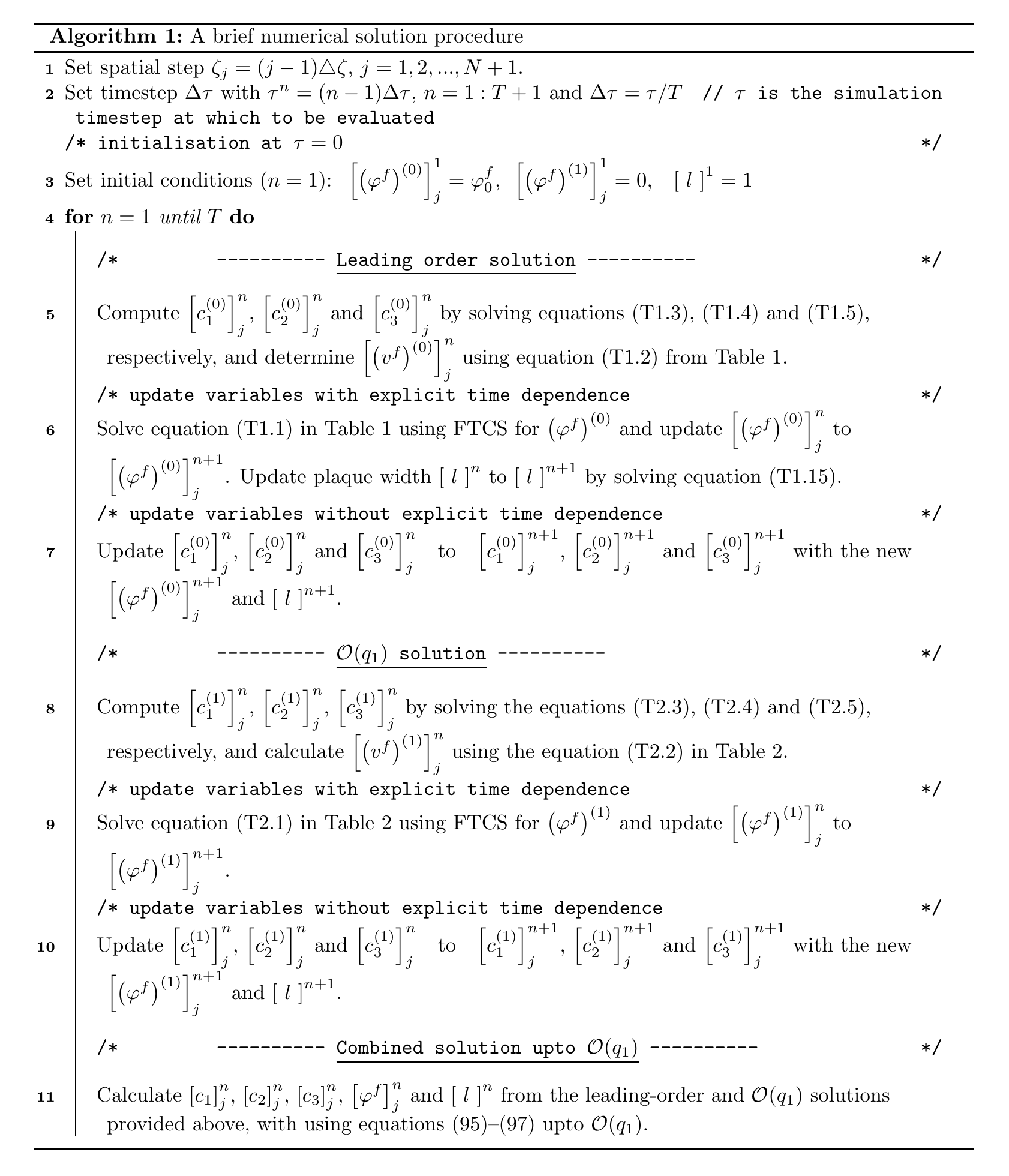}
\label{Algorithm}
\end{figure}

\noindent
The systems in leading order and $\mathcal{O}(q_1)$ are discretized based on the approximations \mbox{(\ref{N1})}-\mbox{(\ref{N5})}. Accordingly, we numerically obtain the solution for both the leading and $\mathcal{O}(q_1)$ problems. Therefore, the combined solution up to $\mathcal{O}(q_1)$ is determined from the equations \mbox{(\ref{P1})}-\mbox{(\ref{P3})}. A brief overview of the numerical procedure for solving the complete system via the algorithm is presented in Algorithm 1. Note that, when applying the central difference \mbox{(\ref{N1})}-\mbox{(\ref{N2})} to the third governing equations from Tables 1 and 2 for $c^{(0)}_{1}$ and $c^{(1)}_1$, respectively, a system of linear equations is obtained for each case. We use the Thomas algorithm to solve these systems. Similar approach is applicable for $c^{(0)}_2$, $c^{(1)}_2$, $c^{(0)}_3$, and $c^{(1)}_3$. However, to meet the CFL condition of the explicit scheme and achieve an accuracy of approximately $10^{-7}$, we took $N = 30$ spatial steps and $10^5$ time steps, ensuring the stability of the corresponding method through the CFL condition. Furthermore, it is important to emphasize that the perturbation expansion method is employed to recast the oxLDL cholesterol transport equation \mbox{\eqref{S27}} into a weakly nonlinear form in $c_1$, rather than a strictly linear form. Nevertheless, the model remains strongly nonlinear in both the leading and $\mathcal{O}(q_1)$ terms, as evidenced by the equations (T1.1) and (T2.1) in Tables \mbox{\ref{tableP1}} and \mbox{\ref{tableP2}} for ${\left(\varphi^{f}\right)}^{(0)}$ and ${\left(\varphi^{f}\right)}^{(1)}$, and their coupling with all other variables. Consequently, numerical methods are employed to solve them in both cases. \\

\section{Numerical Simulations}
\noindent
In this section, we systematically analyze the numerical results on the oxidized LDL cholesterol concentration, intracellular cholesterol concentration, volume of inflammatory phase (foam cells + macrophages), concentration of f-cytokines responsible for macrophage migration, the variation of plaque width over time, etc. First, we set the values of parameters of interest. Subsequently, numerical simulations are performed.
\subsection{\textbf{Model Parameterisations}}
\noindent
The development of plaque in atherosclerosis occurs gradually over time. \citep{badimon1990regression,tangirala1999regression} have indicated that in vivo investigations carried out on rabbits and ApoE-/- or Ldlr-/- mice have demonstrated that the onset of atherosclerotic plaque formation occurs over several days, ranging between $30$ and $90$ days. Therefore, in this study of early plaque development, the simulation time scale is considered to be in days. The clinical study of \citet{grootaert2015defective} mentions an average intimal thickness as $75\mu \text{m}$ in the thoracic aorta of an ApoE-/- mice fed a high-fat diet for 10 or 14 weeks. Since the early plaque growth is restricted within the intima, one can choose the initial plaque width as $l_0=10^{-3}$ cm. As ApoE-/- mice have been found to more closely resemble the human metabolite signature of increased carotid intima-media thickness than other animal models of cardiovascular disease \citep{saulnier2018ldlr}, the choice of $l_0$ is relevant to human atherosclerosis as well.
\begin{figure}[h!]
\centering
\includegraphics[width=0.7\textwidth]{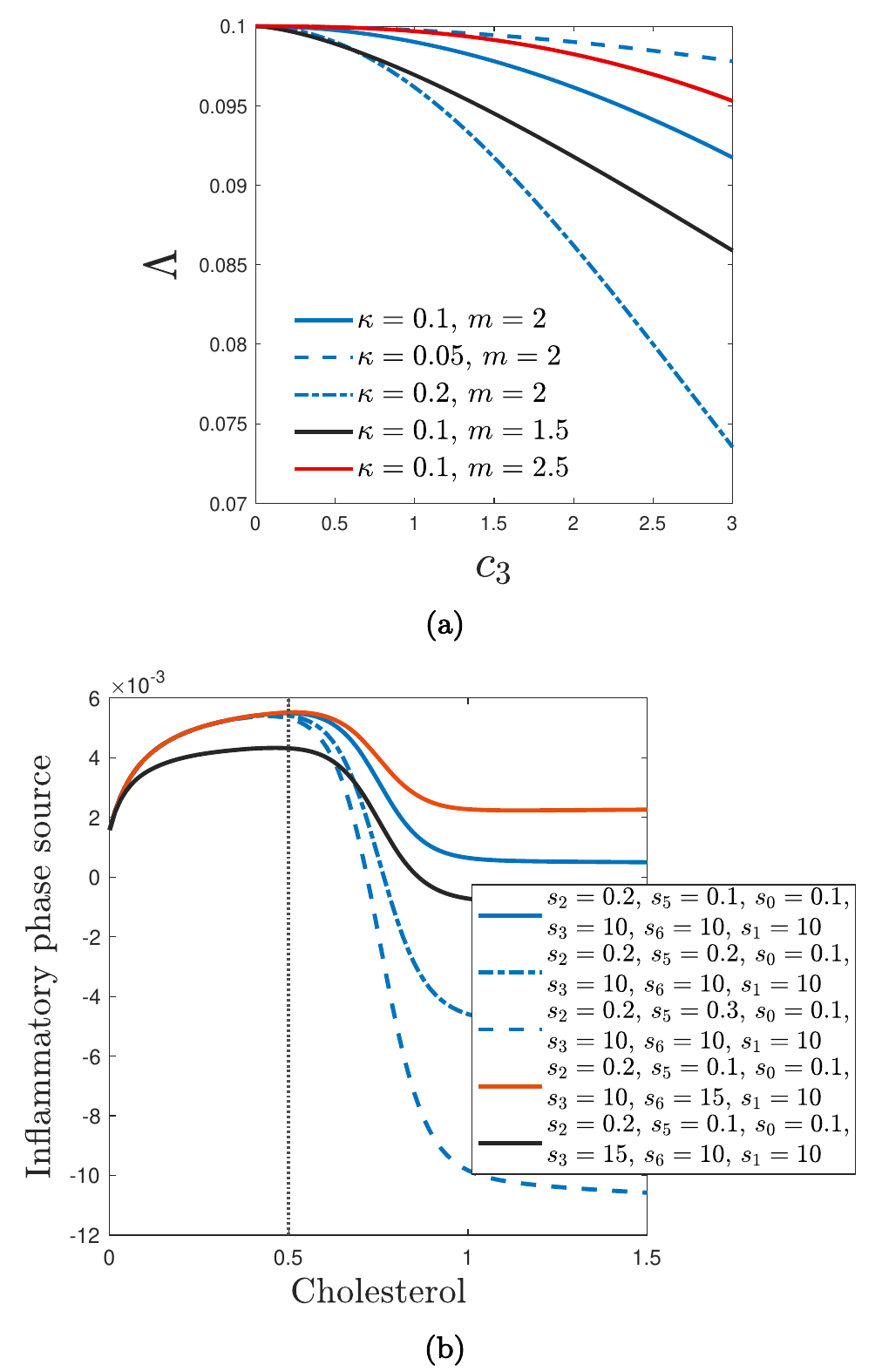}
\caption{Variation of (a) chemotactic function $\Lambda$ \mbox{(\ref{M21})} with respect to $c_3$, (b) inflammatory phase source $\daleth^f$ \mbox{(\ref{M17})} with cholesterol corresponding to various parameters described in the figures. Here, cholesterol primarily signifies oxLDL cholesterol ($c_1$), while intracellular cholesterol ($c_2$) is accounted for using the relation $c_2 = r_c c_1$ to estimate parameters. In (b), the blue curves (solid, dash-dotted, and dashed) correspond to $s_2 > s_5$, $s_2 = s_5$, and $s_2 < s_5$, respectively, while the solid curves (red, blue, and black) indicate $s_3 < s_6$, $s_3 = s_6$, and $s_3 > s_6$, respectively.}
\label{Figure3}
\end{figure}
According to \citet{hao2014ldl}, the diffusion coefficient of LDL or oxidized LDL is $29.89$ $\text{cm}^2 \text{day}^{-1}$. Consequently, the diffusion coefficient for oxLDL cholesterol and intracellular cholesterol can be selected as $29.89$ $\text{cm}^2 \text{day}^{-1}$. In addition, the diffusion coefficient for f-cytokines (particularly for MCP-1) is reported as $17.28$ $\text{cm}^2 \text{day}^{-1}$ \citep{hao2014ldl}. $q_0$ represents the rate of metabolism of oxidized LDL by macrophages, with its value reported as $10^4$ $\text{day}^{-1}$ \citep{mckay2005towards}. Inflammatory cell death leads to the release of intracellular cholesterol into the dead materials. Consequently, the foam cell apoptosis rate can be assumed to correspond to the decay or degradation rate of intracellular cholesterol, represented by the parameter $q_2$. Following \mbox{\citet{ahmed2024hdl}}, the value of $q_2$ is taken as $0.25$ $\text{hr}^{-1}$, or equivalently, $6~\text{day}^{-1}$. In addition, \citet{hao2014ldl} and \citet{chen2012hypoxia} respectively support the production and decay rates of f-cytokines (i.e., $q_4$ and $q_3$) as $0.96 \times 10^{10}$ $\text{g}^{-1} \text{cm}^{3} \text{day}^{-1}$ and $1.73$ $\text{day}^{-1}$. As reported by \citet{hao2014ldl}, within the intima, the concentration of oxLDL and f-cytokines are $7\times 10^{-4}-1.9\times10^{-3}$ $\text{g} \text{cm}^{-3}$ and $3 \times 10^{-10}$ $\text{g} \text{cm}^{-3}$ respectively. The chemotactic sensitivity parameter $\chi$, as mentioned in Eq. (\ref{M21}) can be set as $15.5 \times 10^{-16}$ $\text{g} \text{cm}^{-1} \text{day}^{-2}$ \citep{kim2010interaction}. With the data set mentioned above, we calculate the corresponding non-dimensional values of the parameters $q_0$, $q_2$, $q_3$, $q_4$ and $\chi$ as listed in Table \ref{table1}. Note that we select the value of $q_1$ consistent with the dimensionless value of $q_0$. In addition, this table contains the calculated dimensionless values of constants $\nu_1, \nu_2$ and $\nu_3$ that are present in equations (\ref{S31})-(\ref{S32}) from the studies of \citet{chalmers2015bifurcation,ougrinovskaia2010ode}. Furthermore, the parameter $r_c$ represents the ratio of intracellular cholesterol internalized by macrophages to the extracellular oxLDL cholesterol concentration at the plaque boundary. According to \mbox{\citet{martin2007individual}}, up to $90\%$ of oxLDL particles are taken up by macrophages via scavenger receptors (SR). Consequently, $r_c$ is selected as $0.8$ for the present analysis.\\
\begin{table}[ht]
\caption{Calculated dimensionless values of the parameters from the mentioned literatures.}
\centering
\begin{tabular}{c c c}%
\hline
Parameter & Dimensionless value & Supporting literature \\
\hline
    $q_0$      ~~~~ & ~~~~    $3.3 \times 10^{-1}$ ~~~~ & ~~~~ \citet{mckay2005towards} \\
    $q_1$      ~~~~ & ~~~~    $ 10^{-2}$ ~~~~ & ~~~~  ---               \\
    $q_2$      ~~~~ & ~~~~    $0.2 \times 10^{-4}$ ~~~~ & ~~~~ \mbox{\citet{ahmed2024hdl}} \\
    $q_3$      ~~~~ & ~~~~    $1 \times 10^{-6} $ ~~~~ & ~~~~ \citet{chen2012hypoxia} \\
    $q_4$      ~~~~ & ~~~~    $ 5 \times 10^{-1}$ ~~~~ & ~~~~ \citet{hao2014ldl} \\
    $\chi$     ~~~~ & ~~~~    $1 \times 10^{-1} $ ~~~~ & ~~~~ \citet{kim2010interaction}\\
    $\kappa$   ~~~~ & ~~~~    $1 \times 10^{-1}$ ~~~~ & ~~~~ ---                \\
    $m$        ~~~~ & ~~~~    $2$ ~~~~ & ~~~~           ---                \\
    $l_0$      ~~~~ & ~~~~    $10^{-3}$ ~~~~ & ~~~~ \citet{grootaert2015defective}  \\
    $\varphi^f_{\#}$ ~~~~ & ~~~~    $10^{-2}$ ~~~~ & ~~~~           ---                \\
    $\nu_1$    ~~~~ & ~~~~    $10^{-2} - 10^1$  ~~~~ & ~~~~  \citet{chalmers2015bifurcation}  \\
    $\nu_2$    ~~~~ & ~~~~    $10^{-2} - 10^1$  ~~~~ & ~~~~  \citet{ougrinovskaia2010ode}  \\
    $\nu_3$    ~~~~ & ~~~~    $10^{-3} - 10^{1}$  ~~~~ & ~~~~  \citet{chalmers2015bifurcation}   \\
    $r_c$    ~~~~ & ~~~~    $0.8$  ~~~~ & ~~~~  \mbox{\citet{martin2007individual}}   \\
\hline
\end{tabular}
\label{table1}
\end{table}
\noindent
It has been noted that there is a need for more literature to accurately determine the values of the non-dimensional parameters $\kappa$ and $m$ in the chemotactic function $\Lambda$. As a result, suitable estimates for these parameters would be very handy. In this regard, the behaviour of $\Lambda$ against $c_3$ is shown in Fig. \ref{Figure3}a to estimate $\kappa$ and $m$. Following \citet{hillen2009user}, the estimates of these parameters can be selected within the ranges of $0<\kappa<1$ and $1<m<3$. The chemotactic function generally has a decreasing nature corresponding to all values within the ranges, as shown in Fig. \ref{Figure3}a. $\Lambda$ is plotted for three different values of $\kappa=0.05, 0.1, 0.2$ and $m=1.5,2,2.5$. It is also that for higher values of $\kappa$, the chemotactic function decays faster; however, this decay is slow for large $m$. Therefore, $\kappa=0.1$ and $m=2$ could be a worthy choice.\\

\begin{table}[ht]
\centering
\caption{Estimated values of the non-dimensional parameters contributing in inflammatory phase source term.}\label{table2}
\begin{tabular}{c c c c}%
\hline
Parameter ~~~~~~~~~~~ & ~~~~~~~~~~~ & ~~~~~~~ & ~~~~~~~~~~~ Value \\
\hline
$s_0$ ~~~~~~~~~~~ & ~~~~~~~~~~~ & ~~~~~~~ & ~~~~~~~~~~~ $10^{-1}$    \\
$s_1$ ~~~~~~~~~~~ & ~~~~~~~~~~~ & ~~~~~~~ & ~~~~~~~~~~~ $10$   \\
$s_2$ ~~~~~~~~~~~ & ~~~~~~~~~~~ & ~~~~~~~ & ~~~~~~~~~~~ $2 \times 10^{-1}$  \\
$s_3$ ~~~~~~~~~~~ & ~~~~~~~~~~~ & ~~~~~~~ & ~~~~~~~~~~~ $10$  \\
$s_4$ ~~~~~~~~~~~ & ~~~~~~~~~~~ & ~~~~~~~ & ~~~~~~~~~~~ $10^{-3}$   \\
$s_5$ ~~~~~~~~~~~ & ~~~~~~~~~~~ & ~~~~~~~ & ~~~~~~~~~~~ $10^{-1} - 1.8\times10^{-1}$  \\
$s_6$ ~~~~~~~~~~~ & ~~~~~~~~~~~ & ~~~~~~~ & ~~~~~~~~~~~ $5\times10^{0}-10\times10^{0}$   \\
$s_7$ ~~~~~~~~~~~ & ~~~~~~~~~~~ & ~~~~~~~ & ~~~~~~~~~~~ $10^{-3}$   \\
\hline
\end{tabular}
\end{table}
\noindent
Similarly, we go through the behaviour of the inflammatory phase net source $\daleth^f$ that appears in Eq. (\ref{M17}) to estimate the value of the parameters $s_0, s_1, s_2, s_3, s_4, s_5$ and $s_6$ (as shown in Fig. \ref{Figure3}b). The parameters $s_0$ and $s_1$ are decided inversely proportional to each other, ensuring that the first term (in the production part) of $\daleth^f$, which increases with $c_3$, adequately represents the growth of the inflammatory phase. Likewise, $s_2$ and $s_3$ are selected to be inversely proportional to each other so that the second term (in the production part) of $\daleth^{f}$ with the growth associated with $c_1$, thereby justifying the observed inflammatory phase development. The inflammatory phase concludes into dead material due to the toxicity of intracellular cholesterol, with $c^{*}_{2}$ representing the reference level of cholesterol-induced toxicity. Consequently, the contribution of the death term (i.e. second part in $\daleth^f$) in the inflammatory cell phase source increases with $c_2$ for values higher than $c^{*}_{2}$. Therefore, throughout the study, $s_4<s_5/s_6$ can be assumed. In addition, $s_5$ and $s_6$ may behave inversely towards the increasing nature of the death term. Hence upon the choice of $s_6$ within $[1,~10]$, it may be suitable to consider $s_5$ in $[0.1, 1]$ with the assumption $s_4<s_5/s_6$. Also, the literature suggests that when $s_1=s_3=s_6=0$ (first-order kinetics), the parameters $s_0$, $s_2$ and $s_5$ lies in the range $0.1 \leq s_0, s_3 \leq 10$ \citep{prakash2010criterion,dey2016hydrodynamics,dey2018vivo}. We can consult Fig. \ref{Figure3}b to adjust the parameter values within the specified ranges. Note that the net source term becomes negative when $s_0, s_2 \leq s_5$, as shown by the blue and dash-dotted lines. This incident may be physically unrealistic as the inflammatory phase forms the plaque. Hence, we can consider $s_2>s_5$. Accordingly, we fix $s_2=0.2$ and $s_5\in[0.1,0.2)$ with $s_0=0.1$. In addition, in each of the following cases, $s_3<s_6$, $s_3=s_6$ and $s_3>s_6$, the corresponding scenarios are shown by the solid red, blue and black curves, respectively. Similar arguments are valid for the case $s_3\leq s_6$. For simplicity, one can choose $s_3 = s_6 = 10$ following the study of \citet{breward2002role} on tumour growth, and $s_1=10$, consistent with those for $s_3$ and $s_6$. However, we consider $s_6 \in [5,10]$ while keeping $s_3 = s_1 = 10$, ensuring that the scenario remains physically realistic. The value of the parameter $s_4$ is considered as $10^{-3}$ to satisfy the condition $s_4<s_5/s_6$. Finally, we select the value of the marginal decay parameter $s_7$ as $10^{-3}$ to align with the order of magnitude of the death term caused by toxicity. The above-estimated parameter values are prescribed in Table \ref{table2}.
\subsection{Sensitivity Analysis}
\begin{figure}
\centering
\includegraphics[width=0.6\textwidth]{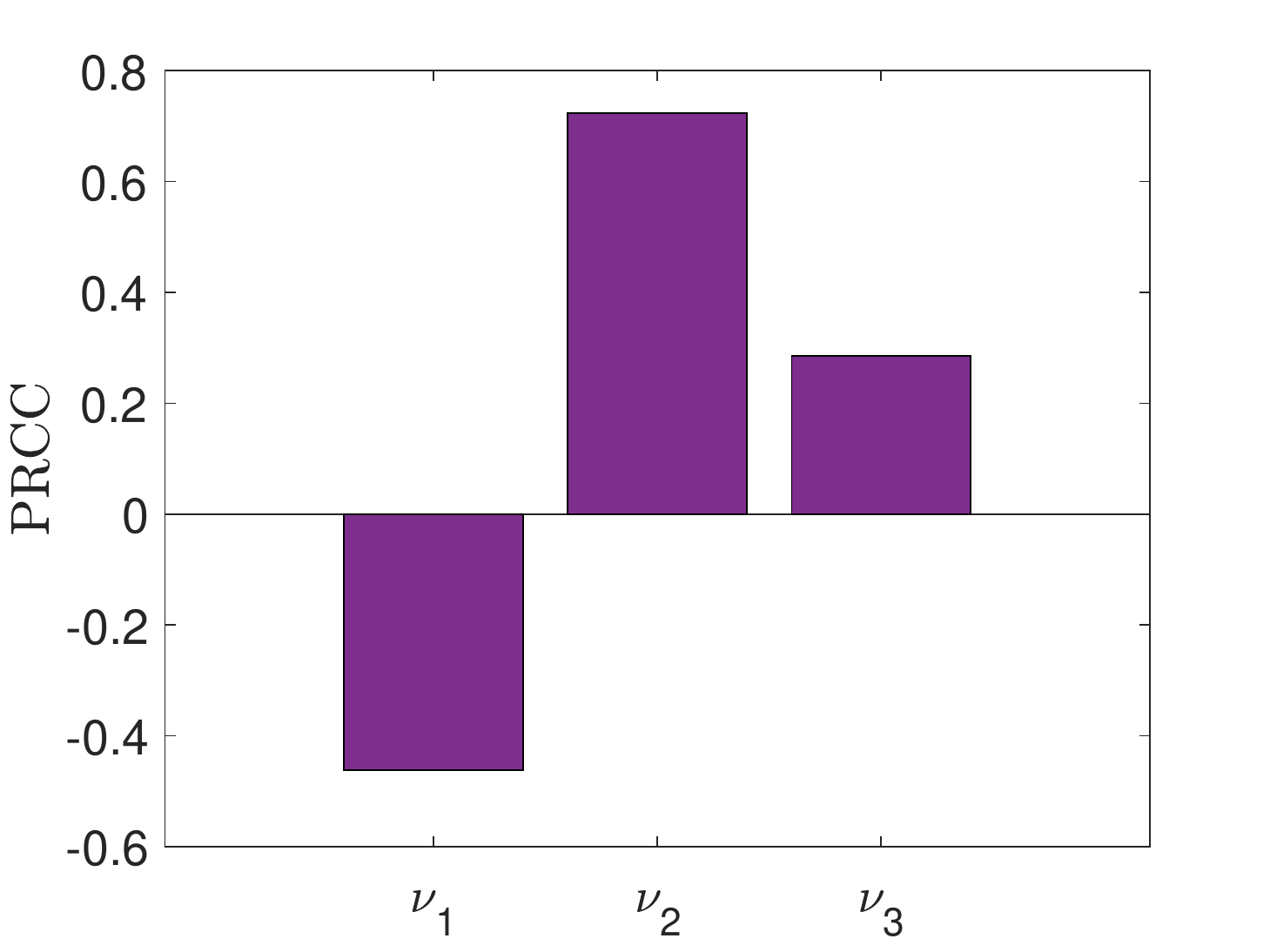}
\caption{The PRCC of flux parameters for sensitivity analysis.}
\label{snsty}
\end{figure}

\noindent
Conducting sensitivity analysis to investigate the impact of various parameters on plaque development in early atherosclerosis improves the reliability of the simulated results. Specifically, we focus on the flux parameters corresponding to oxLDL cholesterol and f-cytokines that appear in the boundary conditions (\ref{M11})-(\ref{M12}). The partial rank correlation coefficient (PRCC) concerning plaque width is calculated for $t=20$ days. The Latin Hypercube Sampling (LHS) is utilized to generate a set of $100$ samples and compute the PRCC values shown in Fig. \ref{snsty}. This methodology follows the sensitivity analysis framework outlined in \citet{marino2008methodology,hao2014ldl}. A positive PRCC indicates a positive correlation, meaning that as the parameter value increases, plaque width also increases. Conversely, a negative PRCC signifies a negative correlation, suggesting that an increase in the parameter value leads to a decrease in plaque width. In this context, $\nu_2$ and $\nu_3$ exhibit a positive PRCC, indicating a positive correlation with plaque width. On the other hand, $\nu_1$ displays a negative PRCC, showing a negative correlation. In other words, $\nu_2$ and $\nu_3$ positively influence plaque development, while parameter $\nu_1$ negatively impacts plaque development. The influence of intracellular-cholesterol-induced toxicity behind the exhibition of such a correlation will be detailed in the next section.
\subsection{\textbf{Results and Discussion}}
\subsubsection{Spatial Patterns in Model Variables}
\begin{figure}[h!]
\centering
\includegraphics[width=\textwidth]{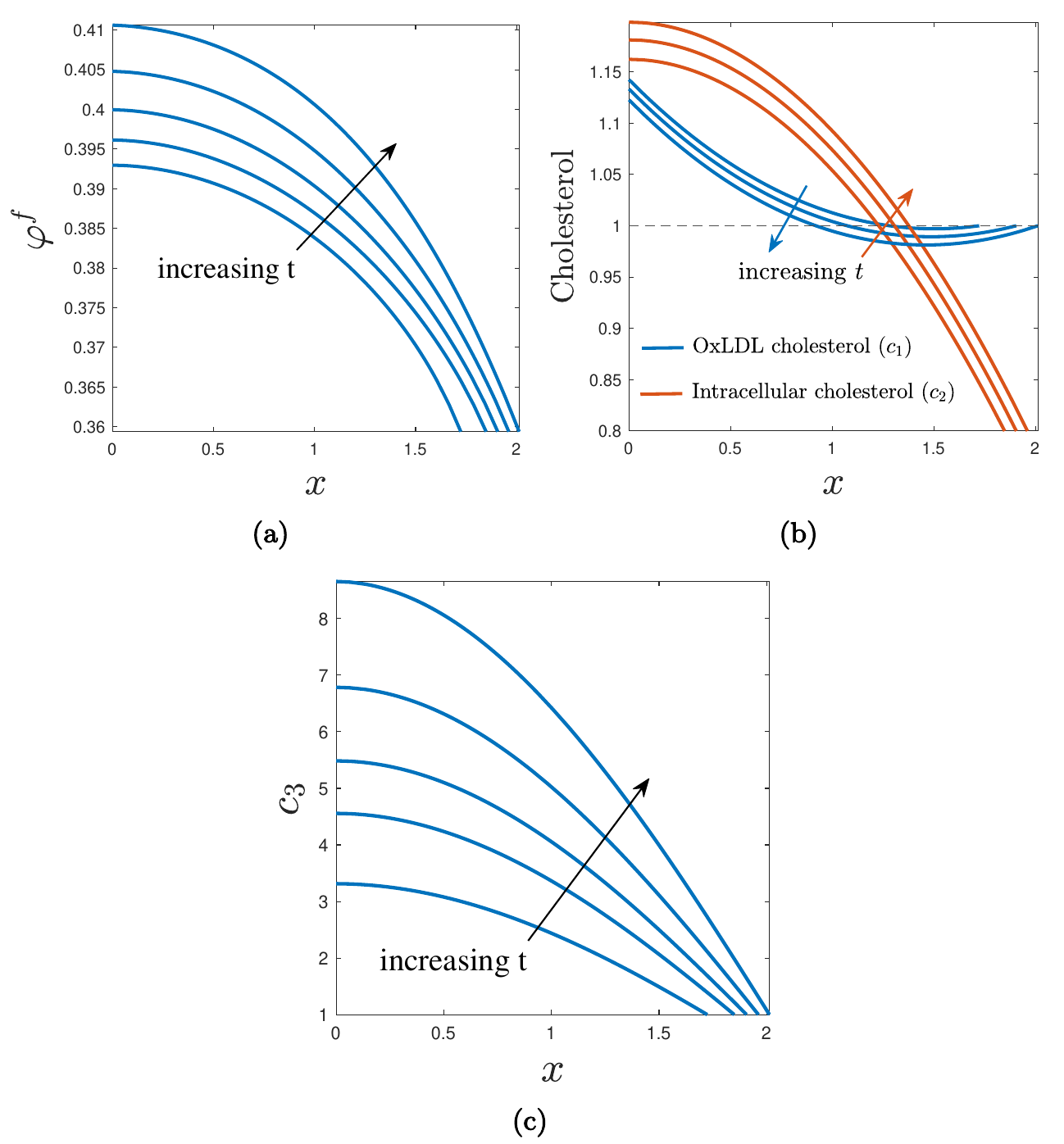}
\caption{(a) Volumetric variation of inflammatory cells, (b) Cholesterol (oxLDL and intracellular) concentration, and (c) f-Cytokines (MCP-1) concentration at every locations within plaque for various $t_{\tiny{\textrm{fix}}}=10,~20,~25,~30$ and $35$ at an early stage of atherosclerosis. The direction of these sequential time points is indicated by the arrows. Note that, in the cholesterol concentration profile, the variations correspond to $t_{\tiny{\textrm{fix}}} = 10,20,\text{and}~30$.}\label{Figure4}
\end{figure}

\noindent
This section discusses the role of the inflammatory cell phase, oxLDL cholesterol, intracellular cholesterol, f-cytokines, and non-inflammatory phase in plaque growth. It provides a thorough analysis of the overall effect of three parameters, namely $\nu_{1}$, $\nu_{2}$, and $\nu_{3}$, on the inflammatory changes that occur in early atherosclerosis and the corresponding impact on plaque development. In addition, the effects of the parameters $s_5$, $s_6$, and $c^{*}_{2}$ on model variables contributing to plaque development are also analyzed. Notably, the inflammatory cell phase predominantly contributes to early plaque expansion. In Fig. \mbox{\ref{Figure4}a,} we observe how the volume of the inflammatory cell phase changes over time at different locations within a growing plaque. The timeline ranges from $t_{\textrm{fix}} = 10$ to $t_{\textrm{fix}} = 35$ and the parameters used are $\varphi^f_{0} = 0.6$, $\varphi^{*} = 0.4$, and $c^{*}_2=0.8$, along with other parameters selected from Tables \mbox{\ref{table1}} and \mbox{\ref{table2}}. The reason for choosing the above range for the timeline is significant and will be emphasized later.
The values of $\varphi^f_{0}$ and $\varphi^{*}$ are initially chosen to include cell-cell interactions, and we continue with these values unless otherwise specified. The volume of the inflammatory cell population is found to be highest at the EII (i.e. $x=0$). Inflammatory cells accumulate mostly towards $x=0$, and their volume decreases towards the moving medial boundary (or IEL boundary), i.e., $x=l(t)$. Note that inflammatory cells are recruited through $x=0$, resulting in a large amount. In due course of time, these cells gradually increase within the intima (see Fig. \mbox{\ref{Figure4}a}). On the other hand, both oxLDL cholesterol and f-cytokines are highly concentrated at the endothelium where they are secreted, and their concentration decreases while moving away from there (to affirm this, see the respective Fig. \mbox{\ref{Figure4}b-\ref{Figure4}c}). Similarly, intracellular cholesterol evolved through the internalization of extracellular oxLDL during consumption. It also shows a high concentration at the endothelium, but decreases away from there (refer to Fig. \mbox{\ref{Figure4}b}). As macrophages consume oxLDL cholesterol and transform into foam cells, the number of inflammatory cells increases, causing reduced oxLDL cholesterol concentration over time. On the other hand, as oxLDL cholesterol is internalized and converted into intracellular cholesterol, the reduction in oxLDL cholesterol leads to a corresponding increase in the intracellular cholesterol concentration over time, as shown in Fig. \mbox{\ref{Figure4}b}. Further, as shown in Fig. \mbox{\ref{Figure4}c}, the concentration of f-cytokines generally increases over time due to their secretion by inflammatory cells.\\

\subsubsection{Dependency of $\varphi^f$ on $c_1$, $c_2$ and $c_3$}
\begin{figure}[h!]
\centering
\includegraphics[width=\textwidth]{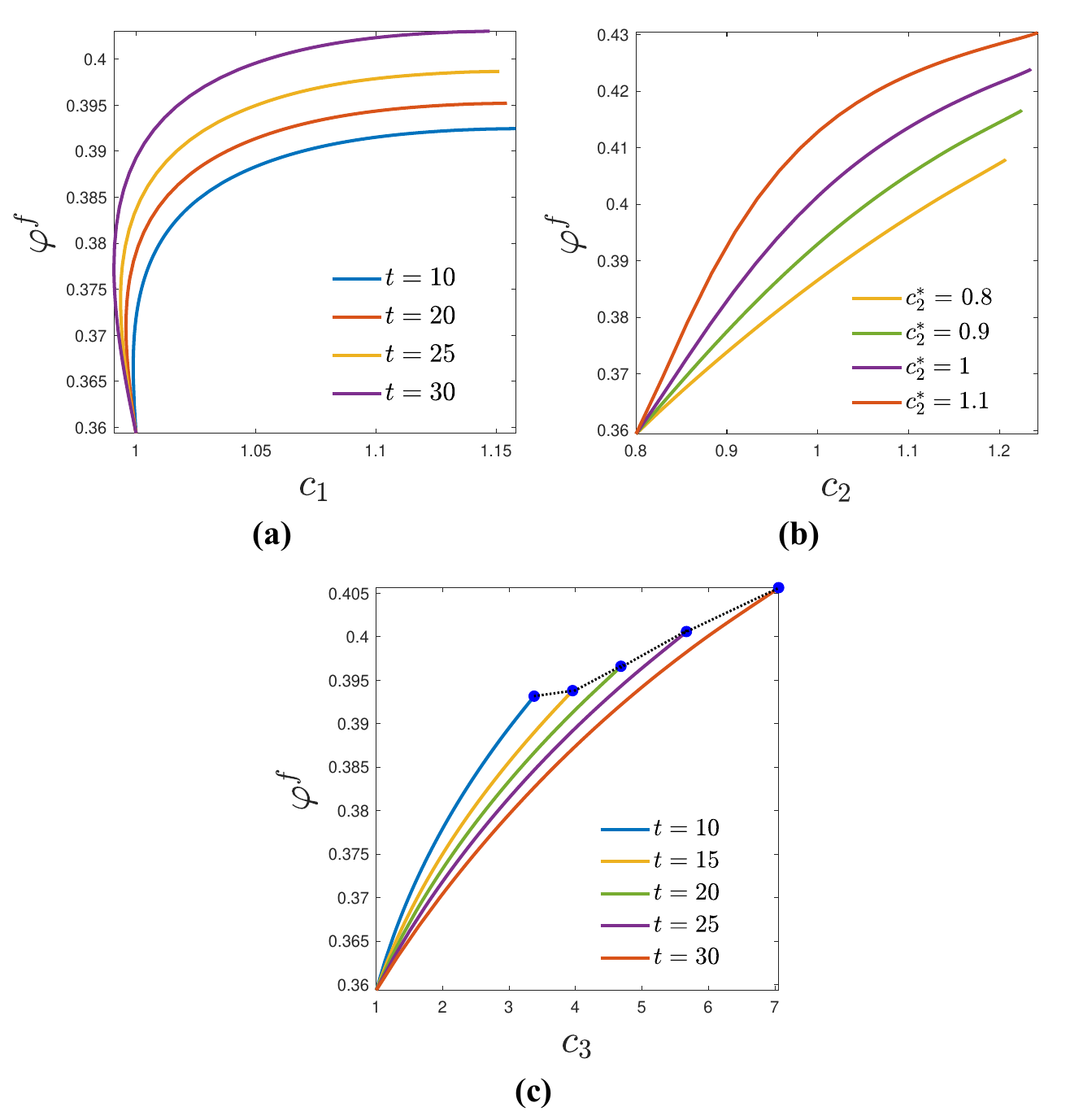}
\caption{Inflammatory cell volume variation with (a) oxLDL cholesterol concentration and (c) f-cytokine concentration at different time points. (b) Variation in volume fraction of inflammatory cells concerning intracellular cholesterol concentration for different toxicity levels $c^{*}_{2}$. The blue dots in (c) represent the maximum volume of inflammatory cells at different times, with the dotted curve indicating its locus.}
\label{Figure5}
\end{figure}

\noindent
Evidently, $\varphi^{f}$, $c_{1}$, $c_{2}$ and $c_{3}$ are mutually dependent. In the intima, macrophages consume oxLDL to become lipid-laden foam cells, and both macrophages and foam cells are integral components of the inflammatory cell phase. Consequently, inflammatory cell volume fraction depends on oxLDL cholesterol concentration. In addition, the volume fraction of inflammatory cells is influenced by intracellular cholesterol concentration. Excessive intracellular cholesterol, beyond the threshold value $c_{2}=c^{*}_{2}$, induces toxicity, leading to inflammatory cell death, as described in Eq. (\mbox{\ref{M17}}). Similarly, the concentration of f-cytokines is influenced by oxLDL cholesterol that stimulates the production of f-cytokines. Fig. \mbox{\ref{Figure5}a} depicts the variation of inflammatory cell volume with oxLDL cholesterol concentration at different times $t$, whereas Fig. \mbox{\ref{Figure5}b} illustrates the variation with intracellular cholesterol concentration for different reference toxicity levels $c^{*}_{2}$. Accordingly, we fix $t=t_{\tiny{\textrm{fix}}}$ where $t_{\tiny{\textrm{fix}}}=10,~20,~25,~30$ etc. Corresponding to each $t_{\tiny{\textrm{fix}}}$, one can identify $\varphi^{f}$ and $c_{1}$ as the function of $x(t=t_{\tiny{\textrm{fix}}})$ such that for each $t=t_{\tiny{\textrm{fix}}}$ we get both the values of $\varphi^{f}$ and $c_{1}$ at each point of the plaque. Similarly, for each $t_{\tiny{\textrm{fix}}}$, the values of $c_{2}$ can be obtained at every plaque point. The inflammatory cell population initially increases with $c_1$ as they are produced when oxLDLs cholesterol particles are metabolized. However, excessive oxLDL uptake leads to the accumulation of high levels of intracellular cholesterol, which induces toxicity and ultimately results in cell death. Consequently, the increase in $\varphi^{f}$ with respect to $c_1$ ceases and tends to meet a nearly constant value (Fig. \mbox{\ref{Figure5}a}). Furthermore, increasing reference toxicity level ($c^{*}_{2}$) induced by the intracellular cholesterol enhances the volume fraction of inflammatory cells. As a result, variation of $\varphi^{f}$ with $c_2$ tends to a stable situation at higher values of $c^{*}_{2}$ (Fig. \mbox{\ref{Figure5}b}). Elevated toxicity levels promote the growth of inflammatory cells due to the absence of toxicity-induced cell death up to that level, resulting in an increased volume fraction of inflammatory cells as $c^{*}_{2}$ rises.\\

\noindent
The concentration of f-cytokines influences the production of inflammatory cells by attracting excess monocytes or macrophages into the intima, where they further transform into foam cells. The corresponding feature is included in our modelling through the boundary condition (\ref{SN9}). In Fig. \ref{Figure5}c, we illustrate the change in inflammatory cell volume fraction against f-cytokines at different times $t_{\tiny{\textrm{fix}}}=10,~15,~20,~25,~30$. With the increasing f-cytokines, the volume fraction of inflammatory cells shows a monotonically increasing behaviour; however, it reaches a maximum value at a particular $c_3$. The corresponding maxima at different times considered are shown with a blue dot. It is observed that the locus of these maxima exhibits a consistently increasing trend, indicating that, over time, the rising volume fraction of inflammatory cells accelerates the plaque growth. The dotted line represents the locus.\\

\subsubsection{Evolution of Plaque Width}
\begin{figure}[h!]
\centering
\includegraphics[width=0.9\textwidth]{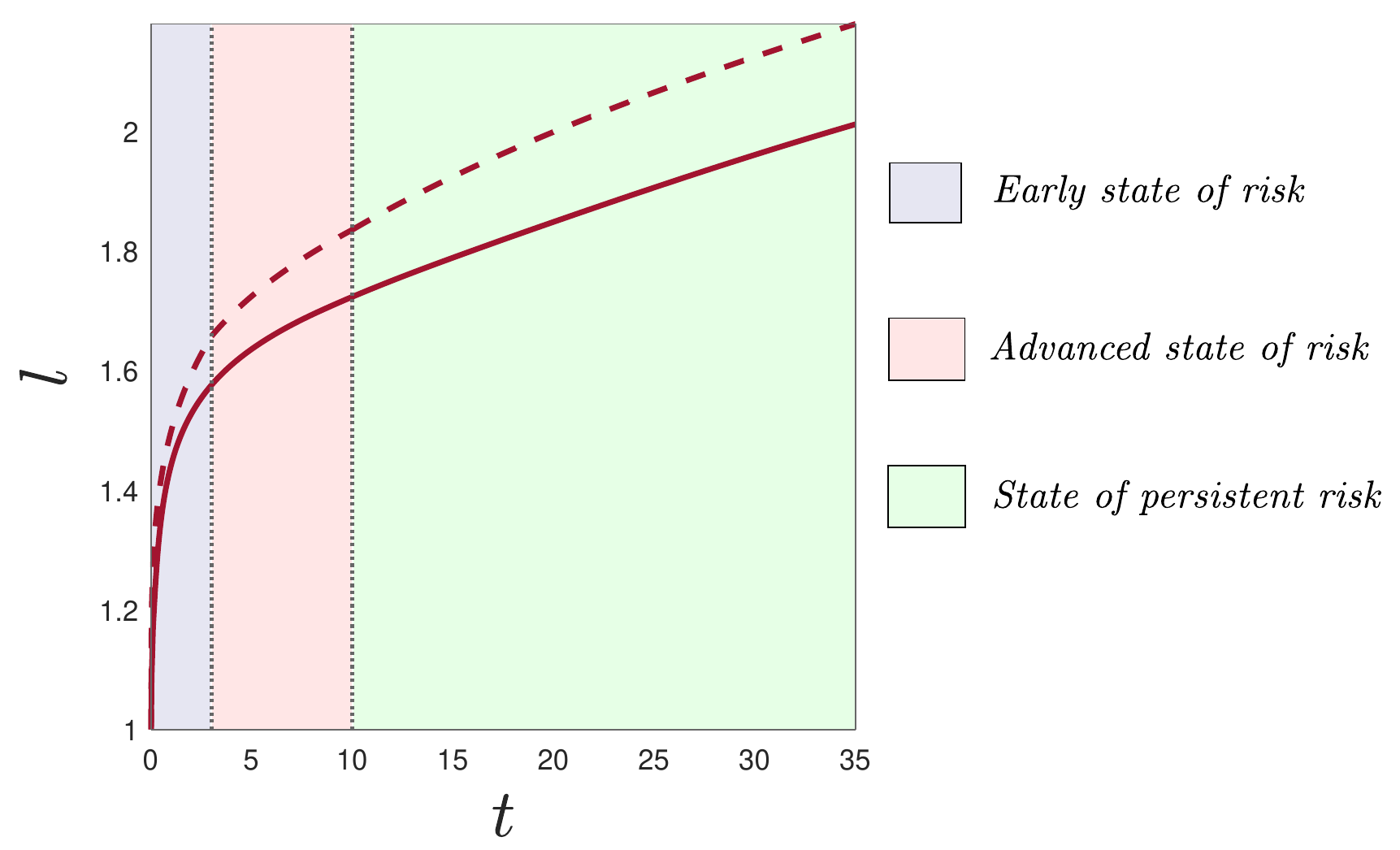}
\caption{Evolution of plaque width $l(t)$ over time $t$. The dashed curve represents the proposed function (\ref{RD1}).}
\label{l_t}
\end{figure}

\noindent
It seems that four factors, namely $\varphi^{f}$, $c_{1}$, $c_{2}$ and $c_{3}$, collectively contribute to the growth of plaque width. Fig. \mbox{\ref{l_t}} displays the change in plaque depth over time (indicated by the solid red line) and shows a rapid initial increase, followed by a gradual decrease in growth rate, which shows no signs of growth saturation. Hence, one would be interested in asking whether the growth rate of plaque width becomes saturated (flattened) or not.
Basically, such flattening depends on the rate constants $s_{5}$ and $s_{6}$ in the inflammatory cell death (due to intracellular cholesterol-dependent toxicity) term in Eq. (\mbox{\ref{M17}}). Later, we discuss this feature in detail. However, at present for fixed $s_{5}$ and $s_{6}$ we are in a position to classify the behaviour of the temporal variation of the plaque width as shown in Fig. \mbox{\ref{l_t}} in three transitions. These are: $0\leq t\leq 3$; $3<t\leq 10$; $t>10$. We then look for correlations in each zone as functions of $\varrho_{1}$, $\varrho_{2}$, $\vartheta$ and $t$. Accordingly, we propose the following
\begin{equation}\label{RD1}
l(t) = \begin{cases}
1 + \frac{1}{2^{\vartheta-2}}t^{\frac{1}{\vartheta+1}}, & \quad 0\leq t\leq3,\\
1 + \varrho_1 + \frac{1}{2^{\vartheta-1}}t^{\frac{1}{\vartheta}},  & \quad 3<t\leq10,\\
1 + \varrho_1 + \varrho_2 + \frac{1}{2^{\vartheta}}t^{\frac{1}{\vartheta-1}},  & \quad t>10.
\end{cases}
\end{equation}
The values of the parameters in the above-proposed function are set as follows: $\vartheta=3$, $\varrho_1 \in [0.29, 0.3]$ and $\varrho_2 \in [0.14, 0.15]$, to ensure that the correlation becomes a reliable approximation of the plaque growth curve. Note that the number of transition zones in the growth curve is the same as the value of $\vartheta$. The dotted line in Fig. \ref{l_t} represents the growth curve corresponding to the above-proposed correlation (\ref{RD1}).\\

\noindent
The similarity in behaviour between the two lines in Fig. \ref{l_t} indicates that Eq. (\ref{RD1}) is a reliable approximation of the solid line, which represents the plaque growth curve in this study. If we closely observe these temporal variations of the plaque growth (either the thick line or the dotted line), we can classify specific zones in terms of the risk that one may have. There is a rapid growth in the \emph{zone 1} ($0\leq t\leq3$), and hence this may be termed as `\textit{early state of risk}'. Consequently, \emph{zone 2} $(3<t\leq10)$ may be termed as `\textit{advanced state of risk}' because by that time, the plaque depth is high enough. Finally, \emph{zone 3} $(t>0)$ may be termed as the `\textit{state of persistent risk}'.
One may observe a linear nature of the plaque growth beyond $t=10$, which implies the patient has a risk of developing complications related to atherosclerosis. In this situation, atherosclerosis is expected to continue for a patient unless it is being treated. The correlations shown for the zone classification and the nomenclature of the nature of risk are the first of their kind in the literature. Based on this classification, more importance may be given to the simulated results beyond $t>10$ days. Accordingly, we choose $\left[10,~35\right]$ days as the focus time range for our study.
One can observe that the zone representing the state of persistent risk exhibits a reduced plaque growth rate as compared to that of other states of risk. The impact of intracellular cholesterol-induced toxicity plays an important role in this regard. Further details in this regard are deferred to the subsequent sections. According to this model, early atherosclerotic plaque grows towards the internal elastic lamina (IEL), which aligns with reality. The growing plaque experiences more rigidity from IEL compared to the damaged endothelium. As a result, in later stages, the ever-increasing plaque has no alternative way except to enter the lumen after causing further damage to the endothelium. This is why atherosclerosis narrows the lumen and prevents the blood flow.
\subsubsection{The Role of oxLDL and f-Cytokines Influxes $\nu_1$, $\nu_2$ and $\nu_3$ on the Plaque Growth Behaviour}
\noindent
To study the role of the influx parameters $\nu_1$ (oxLDL cholesterol), $\nu_2$ (f-cytokines), and $\nu_3$ (f-cytokines) on the model variables $\varphi^{f}$, $c_{1}$, $c_{2}$ and $c_{3}$, it is imperative to express these model variables as functions of either spatial or temporal variables. It is seen that there are local oscillations for each of the variables $\varphi^{f}$, $c_{1}$, $c_{2}$ and $c_{3}$. Hence, we compute the time averaged quantities and analyze their variation. For this purpose, we integrate $\varphi^{f}$, $c_{1}$, $c_{2}$, and $c_{3}$ within the range $[0,~l(t)]$ to obtain the following averaged variables respectively:
\begin{subequations}\label{RD2-4}
\begin{equation}\label{RD2}
\textrm{Averaged volume of inflamatory cells }: \left<\varphi^{f}\right> = \frac{1}{l(t)}\int_{\textrm{x}=0}^{l(t)}\varphi^{f}(\textrm{x},t)\textrm{dx},
\end{equation}
\begin{equation}\label{RD3}
 \textrm{Averaged oxLDL cholesterol concentration}: \left<c_{1}\right> = \frac{1}{l(t)}\int_{\textrm{x}=0}^{l(t)}c_{1}(\textrm{x},t)\textrm{dx},
\end{equation}
\begin{equation}\label{RD3_1}
 \textrm{Averaged intracellular cholesterol concentration}: \left<c_{2}\right> = \frac{1}{l(t)}\int_{\textrm{x}=0}^{l(t)}c_{2}(\textrm{x},t)\textrm{dx},
\end{equation}
and
\begin{equation}\label{RD4}
 \textrm{Averaged f-cytokines concentration}: \left<c_{3}\right> = \frac{1}{l(t)}\int_{\textrm{x}=0}^{l(t)}c_{3}(\textrm{x},t)\textrm{dx}.
\end{equation}
\end{subequations}
Henceforth, rest of the discussion is based on the above-averaged quantities. The integrals are evaluated using standard Newton's cotes formulae. The numerical solutions of $\varphi^{f}$, $c_{1}$, $c_{2}$ and $c_{3}$ are subjected to Simpson's 1/3 rule within the interval $[0,~l(t)]$ for each $t\in[0,~T]$. Note that $\left<\varphi^{f}\right>$, $\left<c_{1}\right>$, $\left<c_{2}\right>$ and $\left<c_{3}\right>$ are functions of time $t$. We choose the time interval of $[10,~35]$ days to analyze further.
\begin{figure}[h!]
\centering
\includegraphics[width=\textwidth]{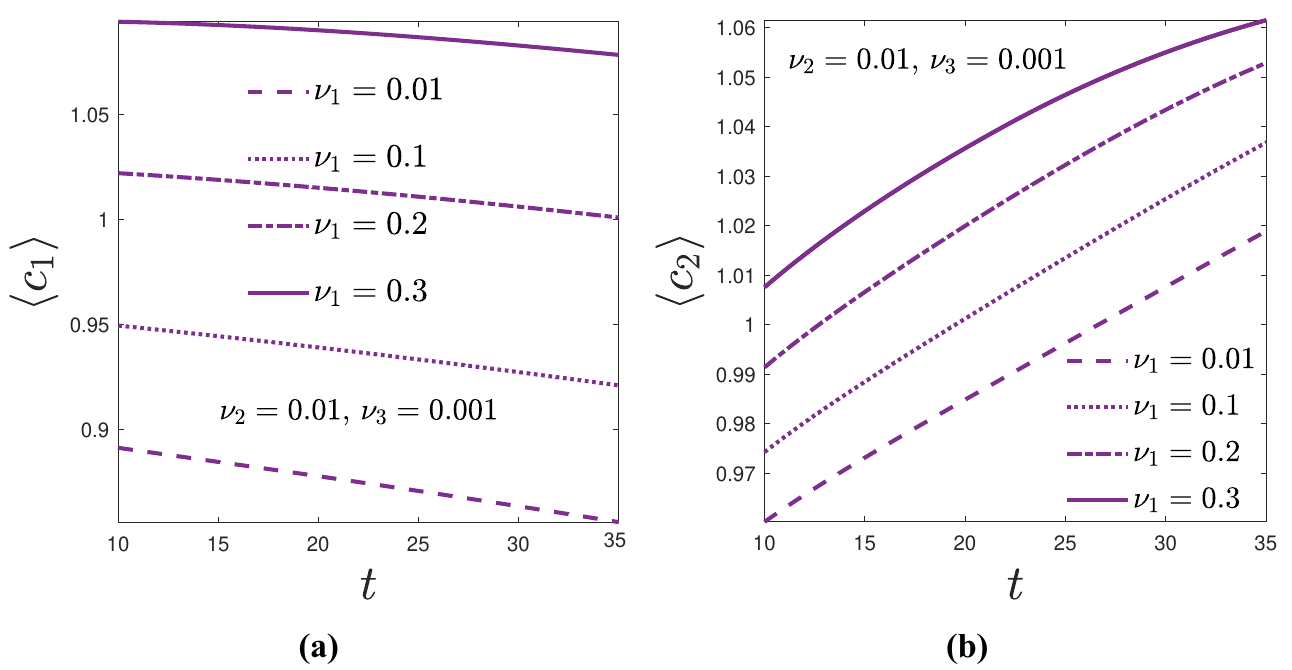}
\caption{Variation of the average concentrations of (a) oxLDL cholesterol and (b) intracellular cholesterol with $t$ for different $\nu_1$ when $\nu_{2}=0.01$ and $\nu_{3}=0.001$.}\label{Figure6}
\end{figure}
Fig. \mbox{\ref{Figure6}a} displays the temporal variation of average oxLDL cholesterol concentration within time domain $[10,~35]$ days corresponding to different influx rates $\nu_1=0.01,~0.1,~0.2,~0.3$. The parameter $\nu_1$ represents a constant oxLDL cholesterol influx rate into the intima through damaged endothelium. According to \mbox{\citet{chalmers2015bifurcation}}, $\nu_1$ can be proportional to the blood LDL cholesterol level. Therefore, the higher magnitude of $\nu_1$ suggests increased concentration of LDL cholesterol in the blood. LDL particles enter the intima through damaged endothelium, where free oxygen radicals oxidize them to become oxLDL. This oxidation process is fast. The timescale for oxidation is much smaller compared to plaque growth. Thus, it is more accurate to assume that oxLDL enters the intima from the lumen through the EII ($x=0$) rather than LDL influx. The concentration of oxLDL cholesterol enhances throughout the intima with the influx parameter $\nu_1$. However, the average oxLDL cholesterol concentration $\left<c_{1}\right>$ eventually reduces. With reducing $\nu_{1}$, a higher rate of decay of $\left<c_{1}\right>$ is realized. The reduced influx rate of oxLDL cholesterol corresponds to when the endothelium layer suffers minor damage. Similarly, a larger influx rate of oxLDL cholesterol can be correlated with a major damaged endothelium. The influx parameter $\nu_1$ positively influences the intracellular cholesterol concentration within the intima, as illustrated in Fig. \mbox{\ref{Figure6}b}. As discussed earlier, oxLDL cholesterol undergoes internalization and is converted into intracellular cholesterol. An increase in $\nu_1$ elevates the concentration of oxLDL cholesterol throughout the intima. Consequently, higher values of $\nu_1$ increase intracellular cholesterol concentrations across the intima. Thus, significant damage to the endothelium may correspond to increased average intracellular cholesterol $\left<c_{2}\right>$. Moreover, $\left<c_{2}\right>$ increases over time, in contrast to $\left<c_{1}\right>$ that decreases over time due to the internalization process.\\

\begin{figure}[h!]
\centering
\includegraphics[width=\textwidth]{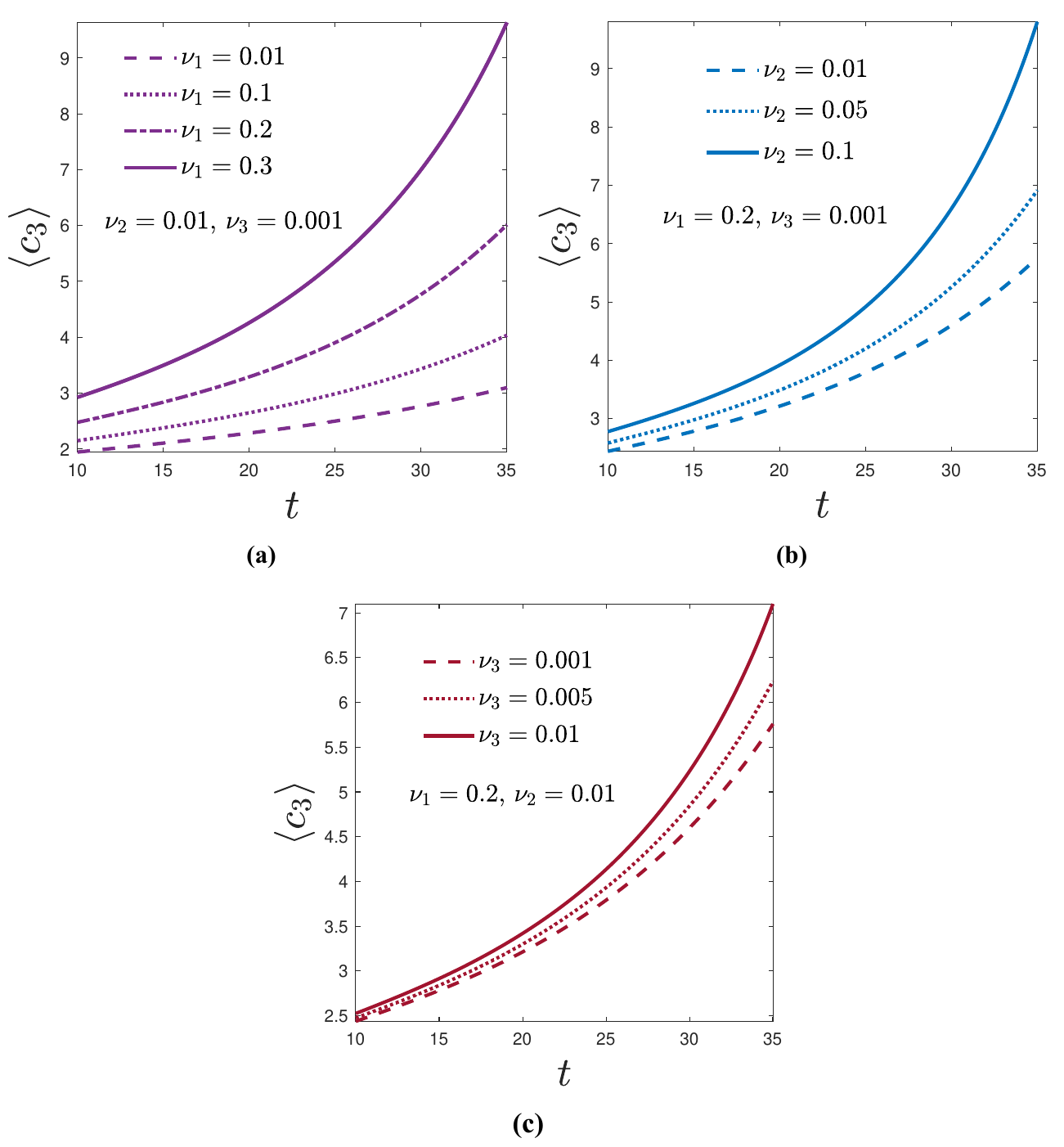}
\caption{The average concentration of f-cytokines versus $t$ for various: (a) $\nu_1$ (when $\nu_{2}=0.01$ and $\nu_{3}=0.001$), (b) $\nu_2$ (when $\nu_{1}=0.2$ and $\nu_{3}=0.001$), and (c) $\nu_3$ (when $\nu_{1}=0.2$ and $\nu_{2}=0.01$).}
\label{Figure6_1}
\end{figure}
\noindent
The influx parameter $\nu_1$ plays an essential role in elevating the concentration of f-cytokines within the intima, as depicted in Fig. \ref{Figure6_1}a. The secretion of f-cytokines at the endothelium is influenced by oxLDL cholesterol, and higher $\nu_1$ enhances the average oxLDL cholesterol concentration within the intima (see Fig. \ref{Figure6}a). Consequently, the concentration of f-cytokines is escalated with higher magnitude of $\nu_1$ due to increased concentration of oxLDL cholesterol. The level of f-cytokines controls the migration of macrophages into the intima. As per the boundary condition (\ref{S14}), the flux of f-cytokines at the EII is directly proportional to oxLDL cholesterol and f-cytokines concentrations ($\left<c_{3}\right>$). Note that f-cytokines are primarily secreted from endothelial cells, and the corresponding flux is regulated by the parameters $\nu_2$ and $\nu_3$. At the EII, the flux of f-cytokines is connected to the oxLDL cholesterol concentration through the parameter $\nu_2$. At the same time, $\nu_3$ regulates the flux of f-cytokines by controlling the corresponding concentration at the interface. Fig. \ref{Figure6_1}b illustrates the variations in the average f-cytokine concentration over time corresponding to $\nu_{2}=0.01,~0.05,~0.1$ when $\nu_{1}=0.2$ and $\nu_{3}=0.001$. Larger values of $\nu_2$ result in a higher influx of f-cytokines at the EII. Consequently, it enhances the average f-cytokine concentration within the intima, as shown in Fig. \ref{Figure6_1}b. Similarly, the more significant value of the influx parameter $\nu_3$ escalates the influx of f-cytokines at the EII. As a result, the average concentration of f-cytokines ($\left<c_{3}\right>$) within the intima rises with higher $\nu_3$, as evidenced in Fig. \ref{Figure6_1}c.\\

\begin{figure}[h!]
\centering
\includegraphics[width=\textwidth]{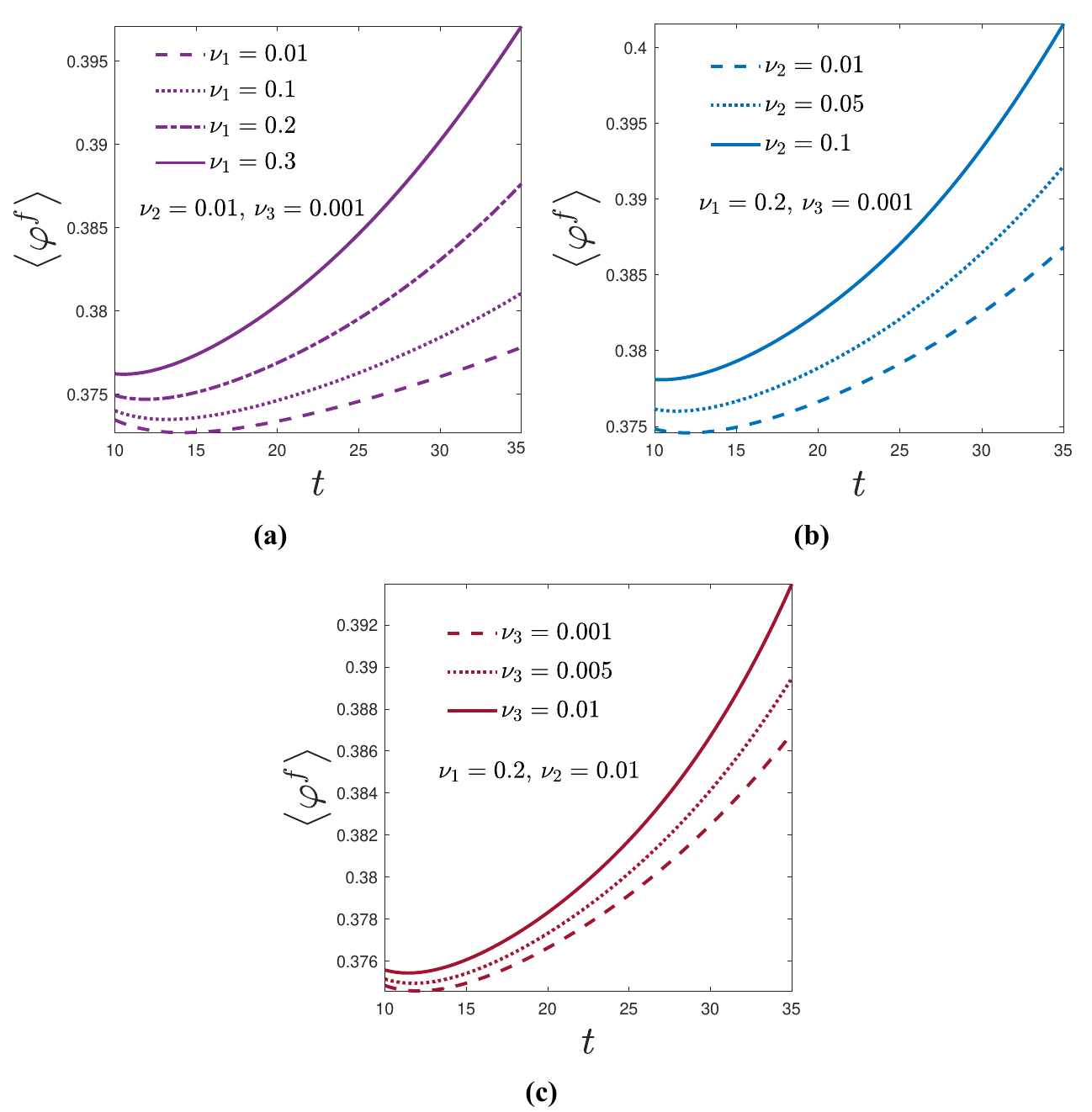}
\caption{Variation of inflammatory cell average volume fraction for different (a) $\nu_1$, (b) $\nu_2$, and (c) $\nu_3$.}
\label{Figure7}
\end{figure}
\noindent
It is worth noting that inflammatory cells primarily contribute to early plaque development. Therefore, we must emphasize the impact of influx parameters $\nu_1$, $\nu_2$, and $\nu_3$ on the average volume of inflammatory cells $\left<\varphi^{f}\right>$. In this regard, Fig. \ref{Figure7}a-\ref{Figure7}c respectively depict the decisive impact of parameters $\nu_1$, $\nu_2$, and $\nu_3$ on $\left<\varphi^{f}\right>$.
In all these three cases, the average volume of inflammatory cells grows with $t$ within the interval $[10,~35]$. $\nu_{1}$, corresponding to the influx of oxLDL cholesterol, becomes responsible for inflammatory cell production. Higher $\nu_1$ correlates with increased concentrations of oxLDL cholesterol, triggering inflammatory response through pro-inflammatory f-cytokines. This process draws monocytes (macrophages within the intima) from the bloodstream to the intima, where they further transform into foam cells by consuming oxLDL. Consequently, one can anticipate a higher production of inflammatory cells. Fig. \ref{Figure7}a supports this phenomenon. The other two influx parameters, $\nu_{2}$ and $\nu_{3}$, are found to exhibit a positive response on the growth of inflammatory cell volume ($\left<\varphi^{f}\right>$) throughout the intima as evidenced in Fig. \ref{Figure7}b and \ref{Figure7}c, respectively. Fig. \ref{Figure7}b illustrates the variation of $\nu_{2}$ in the averaged inflammatory cell volume fraction, while Fig. \ref{Figure7}c displays the variation of $\nu_3$. It can be observed that both $\nu_{2}$ and $\nu_{3}$ contribute to the inflammatory cell volume fraction $(\left<\varphi^{f}\right>)$ with their respective elevations. The positive role of these influx parameters in influencing the average f-cytokines concentration within the intima has already been investigated. Consequently, the recruitment of macrophages from the bloodstream and, subsequently, the production of new inflammatory cells increase.\\

\begin{figure}[h!]
\centering
\includegraphics[width=\textwidth]{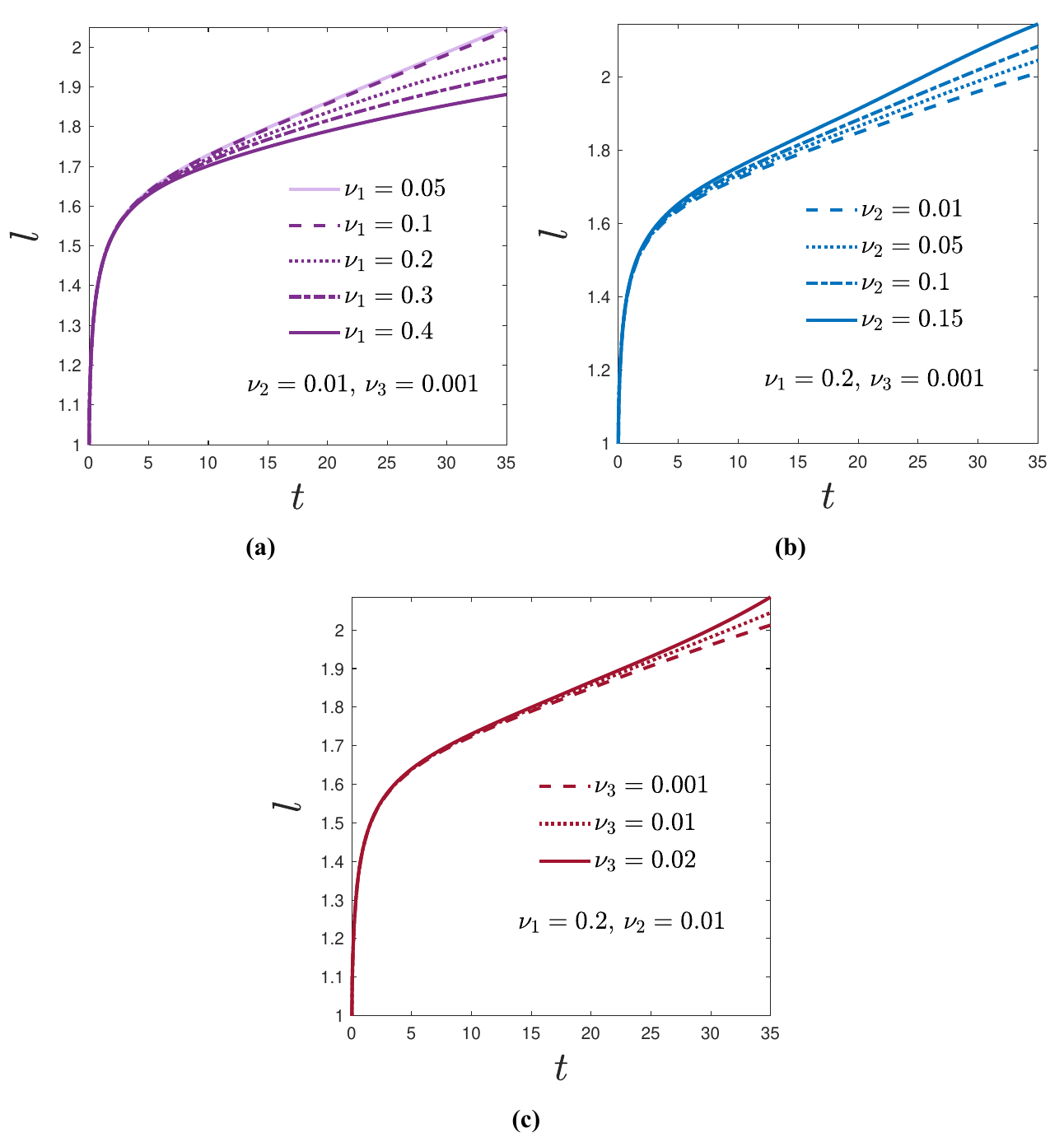}
\caption{Plaque width for various (a) $\nu_1$, (b) $\nu_2$, and (c) $\nu_3$.}
\label{Figure8}
\end{figure}

\noindent
As noted earlier, early plaque development primarily depends on the inflammatory cell phase in the presence of oxLDL cholesterol and an active f-cytokine gradient. Consequently, the plaque depth is modulated by $\nu_1$, $\nu_2$, and $\nu_3$, as illustrated in Figs. \mbox{\ref{Figure8}a}-\mbox{\ref{Figure8}c}, respectively. We have already classified plaque growth and its impact on early atherosclerosis into three regimes: early state, advanced state, and persistent state of risk. These three parameters mainly affect the third regime. Hence, one can comprehend the long-term impact of these parameters.
Fig. \mbox{\ref{Figure8}a} illustrates that the plaque growth rate reduces with increasing values of $\nu_1$ at any given time. This reduction is more significant over time, suggesting a potential long-term effect of $\nu_1$ on plaque formation. Higher $\nu_1~(>0.1)$ leads to a significant increase in oxLDL cholesterol concentration within the intima, inevitably resulting in a substantial buildup of foam cells within this layer. Higher oxLDL influx leads to a considerable accumulation of intracellular cholesterol upon ingestion, which later on is responsible for toxicity-based inflammatory cell death. As plaque growth primarily depends on the inflammatory tissues, the production of inflammatory cells decreases at larger $\nu_1$, indicating a reduction in the plaque growth rate. On the other hand, with a low influx rate of oxLDL cholesterol, the concentration of intracellular cholesterol reduces (Fig. \mbox{\ref{Figure6}b}).  Consequently, the impact of induced toxicity is less over the inflammatory cell population. When the oxLDL influx rate $\nu_{1}\leq0.1$, a marginal change in the plaque depth occurs for a further smaller magnitude of $\nu_{1}$. In this situation, the inflammatory cells deposited within the intima (though not in large numbers) do not significantly encounter toxicity death. As a result, plaque depth profile increases sharply over time due to its resistance to cholesterol-induced toxicity. Now, the regime of persistent risk has become more expansive. It is important to note that plaque growth rate tends to slow down unless sufficient dead materials are produced from the inflammatory cells. However, this aspect is outside the scope of the current study. As a result, plaque growth is delayed in advanced stages (increased oxLDL influx rate). The present study shows intracellular cholesterol-induced toxicity is critical to stabilising plaque growth.\\

\noindent
The role of influx parameters $\nu_2$ and $\nu_{3}$ corresponding to f-cytokines over the plaque growth is found to behave opposite to $\nu_1$ (see Fig. \ref{Figure8}b and \ref{Figure8}c). A notable impact of f-cytokines within intima is experienced at larger values of $\nu_2$ and $\nu_3$ that consequences a high concentration of f-cytokines. Consequently, the inflammatory response causes more macrophages to migrate to the intima under the influence of increased f-cytokines, and causes increased foam cell population. Hence, further recruitment and proliferation of inflammatory cells occur within the intima. This elevates the growth rate of the plaque width. Consequently, the growth in early plaque width is positively correlated with the values of $\nu_2$ (Fig. \ref{Figure8}b) and $\nu_3$ (Fig. \ref{Figure8}c). The comprehensive analysis affirms that the oxLDL cholesterol influx parameter $\nu_1$ positively influences the flattening of the plaque growth profile over time, primarily due to intracellular cholesterol-induced toxicity. On the other hand, the influx parameters $\nu_2$ and $\nu_3$, associated with f-cytokines, disrupt this flattened behaviour.
\begin{figure}[h!]
\centering
\includegraphics[width=\textwidth]{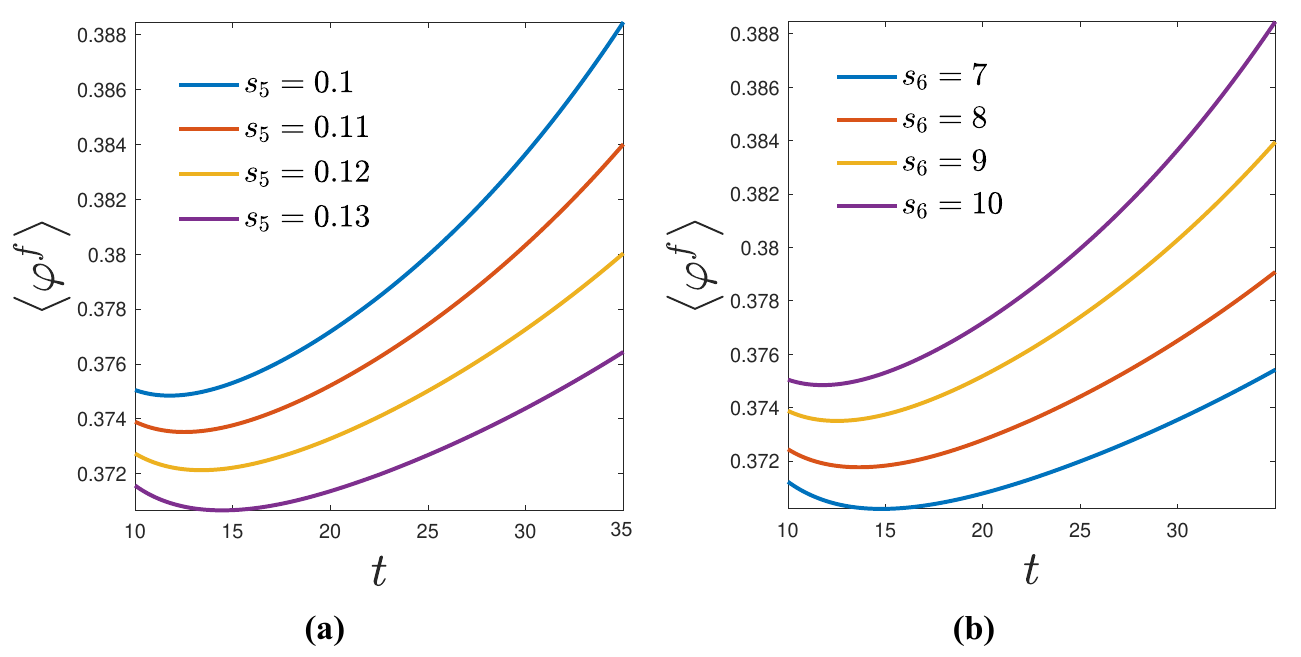}
\caption{Variation of average inflammatory cell volume fraction for different (a) $s_5$, and (b) $s_6$.}
\label{Figure11}
\end{figure}
\subsubsection{The Influence of Parameters $s_5$ and $s_6$ on Plaque Development}
\noindent
The kinetics parameters $s_5$ and $s_6$ present in Eq. \mbox{\eqref{M17}} correspond to the death of inflammatory cells caused by intracellular cholesterol-induced toxicity. The death term in the inflammatory cell phase source ($\daleth^f$) increases with $c_2$ for values exceeding $c_2^{*}$. Accordingly, the condition $(s_4<s_5/s_6)$ holds. The effect of $s_5$ and $s_6$ on the average volume fraction of inflammatory cells $\left<\varphi^{f}\right>$ is presented in Figs.~\mbox{\ref{Figure11}a} and \mbox{\ref{Figure11}b}, respectively. In both cases, the variation in the average volume of inflammatory cells with $t$ is shown within the interval $[10,35]$, corresponding to the persistent state of the risk regime, to elucidate the long-term impact.  Increasing values of the parameter $s_5$ enhances the toxicity-induced death for each fixed $c_2 (> c_2^{*})$.
As a result, an increase in $s_5$ accelerates the death of inflammatory cells due to toxicity, thereby reducing their volume as shown in Fig.~\mbox{\ref{Figure11}a}. In contrast, the parameter $s_6$ is found to exert a positive influence on $\left<\varphi^{f}\right>$ throughout the intima, as demonstrated in Fig.~\mbox{\ref{Figure11}b}. Increasing values of $s_6$ reduces the toxicity-induced death, thereby diminishing the death of inflammatory cells caused by toxicity. This reduction enhances the volume of inflammatory cellsas shown in Fig.~\mbox{\ref{Figure11}b}.\\

\begin{figure}[h!]
\centering
\includegraphics[width=\textwidth]{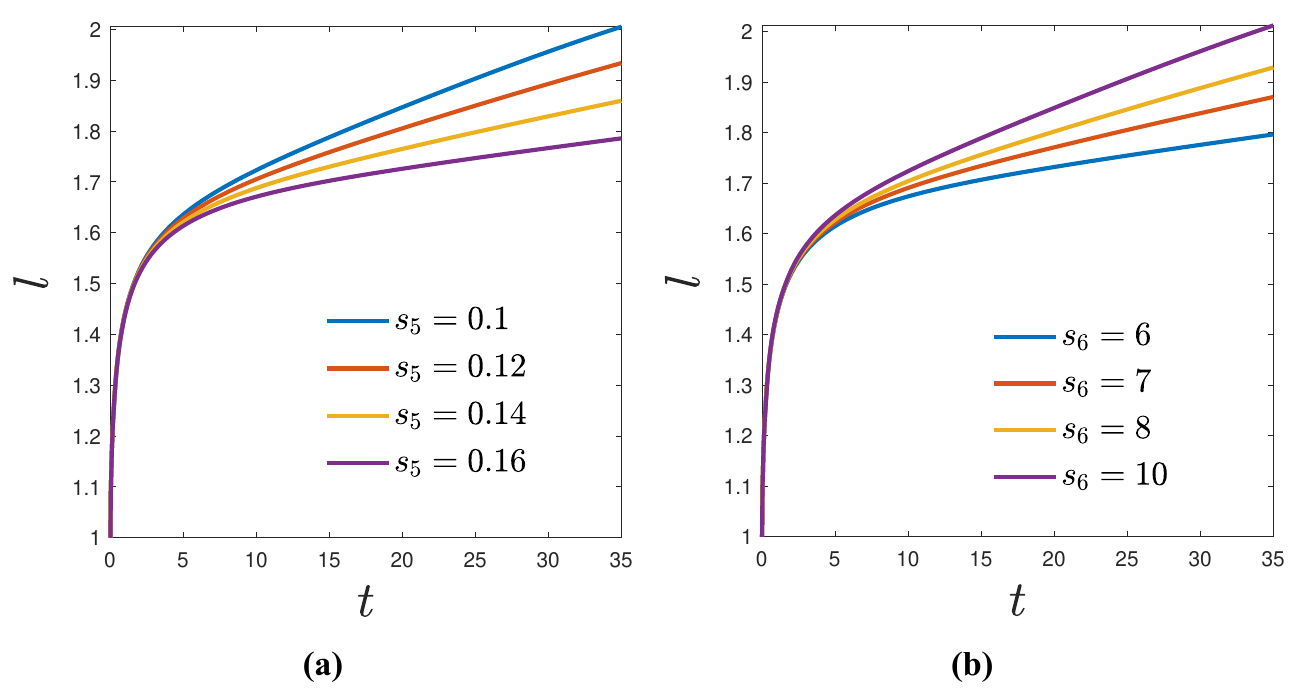}
\caption{The variation in plaque width growth for various (a) $s_5$, and (b) $s_6$.}
\label{Figure12}
\end{figure}

\noindent
As detailed before, early plaque development is highly influenced by the toxicity-based death of inflammatory cells. $s_5$ and $s_6$ are two parameters that regulate this death. Consequently, these two parameters are bound to affect plaque depth with their effects illustrated in Figs.~\mbox{\ref{Figure12}a}-\mbox{\ref{Figure12}b}, respectively. These parameters predominantly affect the persistent risk state, exhibiting long-term impact. A slow increase (a reduction in the plaque growth rate) in the plaque depth is resulted for higher values of $s_5$ (Fig.~\mbox{\ref{Figure12}a}). $s_5$ has a potential long-term effect on the death rate of inflammatory cells due to intracellular cholesterol-induced toxicity. On the other hand, increased $s_6$ reduces the toxicity-induced death rate of inflammatory cells, and enhances the plaque depth as illustrated in Fig.~\mbox{\ref{Figure12}b}. Therefore, the parameters $s_5$ and $s_6$, associated with cholesterol toxicity-induced cell death, play a significant role in stabilizing plaque growth. Consequently, the plaque growth profile shows a plateau for higher $s_5$ or lower $s_6$. Elevated intracellular cholesterol slows down plaque growth, promoting a more stable growth pattern. This insight underscores the significance of cholesterol management in maintaining vascular health.
\begin{figure}[h!]
\centering
\includegraphics[width=\textwidth]{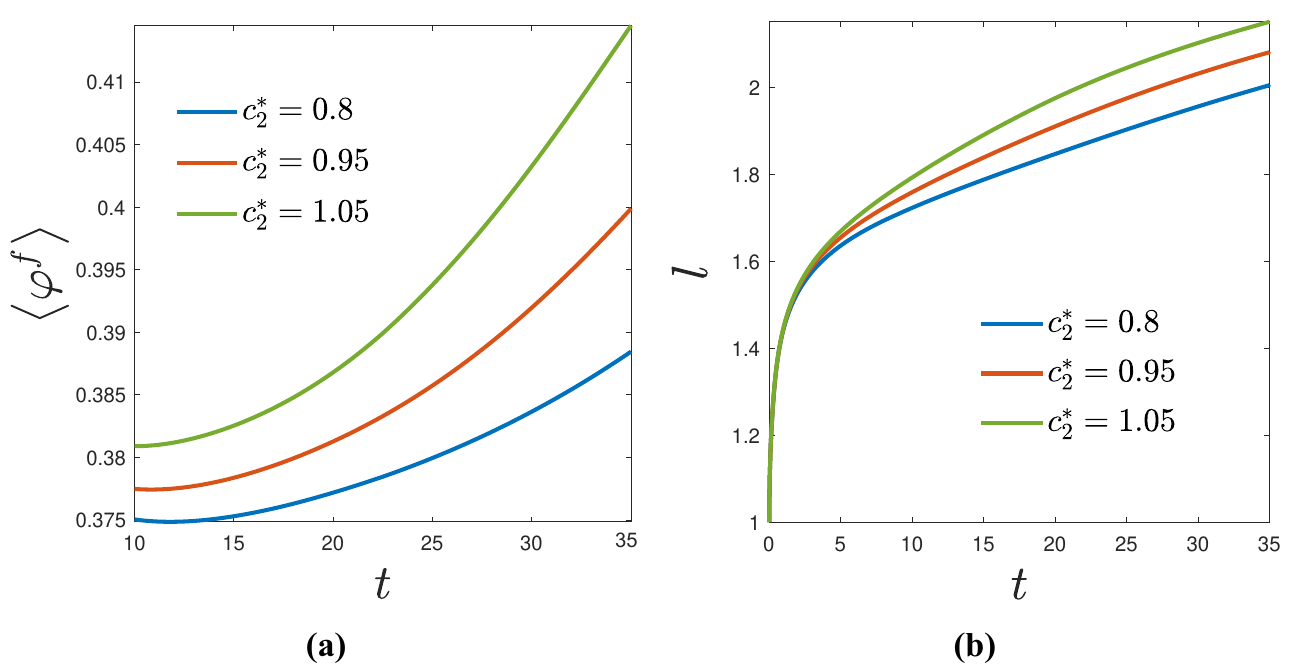}
\caption{(a) Distributions of inflammatory cell average volume fraction, and (b) plaque width variation with $t$ within the intima corresponding to various $c^{*}_2$.}
\label{Figure9}
\end{figure}
\subsubsection{Varied Toxicity Level Induced by Intracellular Cholesterol}
\noindent
As discussed earlier, the toxicity caused by intracellular cholesterol leads to inflammatory cell death, which contributes significantly to plaque formation. The impact of different toxicity levels, denoted by $c^{*}_{2}$, on the growth of inflammatory cell volume and plaque width is shown in Figs. \mbox{\ref{Figure9}a}-\mbox{\ref{Figure9}b} respectively. Inflammatory cells suffer death when the cholesterol reaches its toxicity level. Higher levels of toxicity result in increased growth of inflammatory cells due to their rapid proliferation and absence of cell death at that level of toxicity. This incident leads to an increased volume fraction of inflammatory cells as $c^{*}_{2}$ rises (Fig. \mbox{\ref{Figure9}a}). Moreover, as depicted in Fig. \mbox{\ref{Figure9}b}, a rise in $c^{*}_{2}$ corresponds to an increased plaque width. A higher $c^{*}_{2}$ leads to an increased population of inflammatory cells, accelerating the development of the plaque formed by their deposition.
\subsection{\textbf{Qualitative Agreement with the Clinical Studies}}
\noindent
We attempt to compare specific results from this study with the existing clinical studies done by \citet{takahashi1995induction} and \citet{stangeby2002coupled}. In the study by \citet{stangeby2002coupled}, LDL concentration within the arterial wall is computed under hypertensive conditions using experimental data from \citet{meyer1996effects}, which involves experiments on a rabbit thoracic aorta. Accordingly, the LDL concentrations within the arterial intima are picked up from the study of \citet{stangeby2002coupled} corresponding to transmural pressures of $p=70, 120, 160$ mmHg. In the in vitro study conducted by \citet{takahashi1995induction}, MCP-1 concentration was determined in the conditioned media from cocultures of human peripheral blood monocytes and human umbilical vein endothelial cells (HUVECs).
\begin{figure}[h!]
\centering
\includegraphics[width=\textwidth]{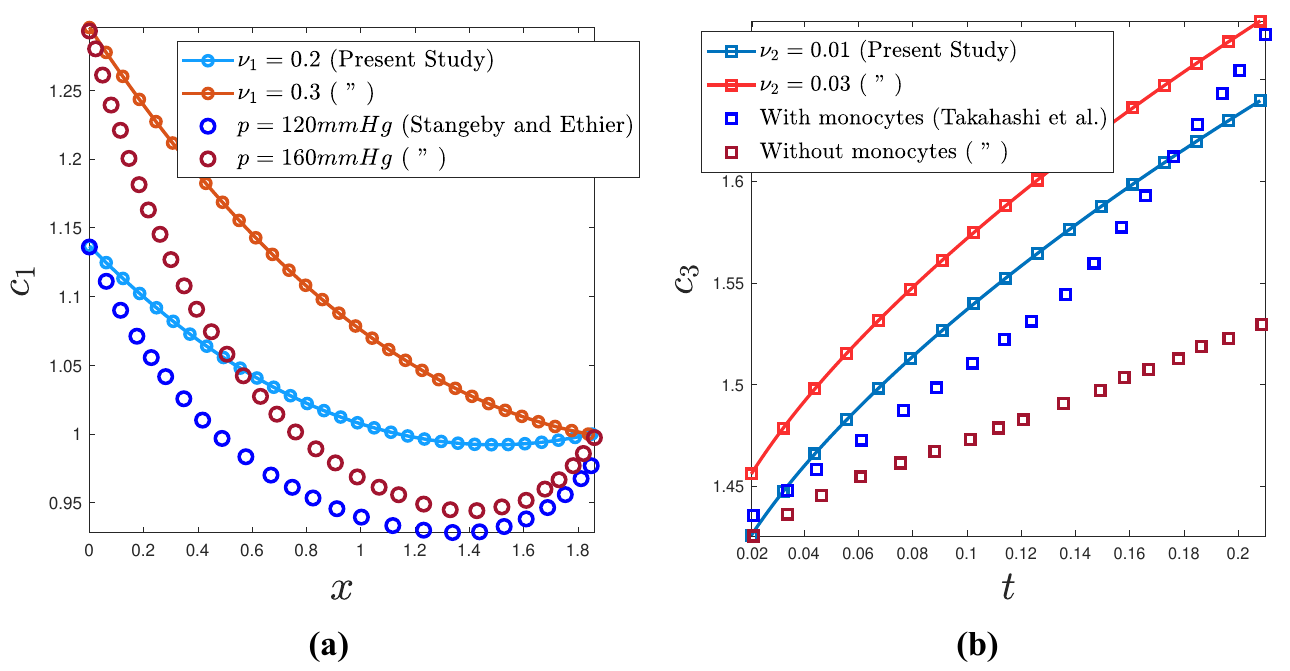}
\caption{Comparison between the present study and the clinical studies done by (a) \citet{stangeby2002coupled} and (b) \citet{takahashi1995induction} along the variation of oxLDL cholesterol and f-cytokines concentrations, respectively. In (a), the variations correspond to different $\nu_1$ for various pressures, and in (b), the variations pertain to different $\nu_2$ for both cases of human umbilical vein endothelial cells (HUVECs) with and without monocytes.}\label{Figure10}
\end{figure}
Correspondingly, MCP-1 concentrations are computed from this study for both conditions: HUVECs with and without monocytes in the presence of interleukin-1 beta (IL-1$\beta$) over an incubation period of 5 hours. In the present study, the diffusible species of f-cytokines include MCP-1 and other cytokines. Consequently, to compare the clinical MCP-1 levels with the current f-cytokines concentration, it is reasonable to consider the effect of pretreatment with IL-1$\beta$. The LDL and MCP-1 concentration data are extracted using the `Grabit' toolbox of MATLAB R2021a. These data are normalized first before comparing with the oxLDL cholesterol and f-cytokines concentrations obtained from the present study. The normalization process is done so that the domain of clinical data maps onto that of the present study through a suitable transformation. Such transformation pivots more on the qualitative agreement rather than the quantitative. In transforming the MCP-1 concentration data, the hour time scale from the clinical study has been converted to the day time scale used in our study, with the initial time for comparison considered as $1/2$ hours (or $0.0208$ days). Note that the present study is focused on oxLDL rather than LDL, as oxidization takes place in a short time while growth is a long-scale process.  \\

\noindent
Figs. \ref{Figure10}a and \ref{Figure10}b delineate the above-mentioned comparative study. In the former, the concentrations of oxLDL cholesterol (this study) and LDL (\citet{stangeby2002coupled}) are shown for $\nu_1=0.2, 0.3$ and $p= 120, 160$, respectively. The latter compares the concentration of f-cytokines (present study) and MCP-1 \citep{takahashi1995induction}, corresponding to $\nu_2 = 0.01, 0.03$, in the cases of HUVECs (Human umbilical vein endothelial cells) with and without monocytes. A good qualitative agreement between the clinical and present studies has been observed in both figures. The comparison in Fig. \ref{Figure10}a shows that the normalized LDL (obtained from \citet{stangeby2002coupled}) and oxLDL cholesterol (obtained from the current study) concentrations behave the same. Both are highly concentrated near the lumen or EII and decrease toward the media or arterial wall. Similarly, the comparison in Fig. \ref{Figure10}b indicates that MCP-1 (obtained from \citet{takahashi1995induction}) and f-cytokines (obtained from the present study) concentrations manifest similar behavior. Specifically, MCP-1 in the clinical study corresponding to the case of HUVECs coincubated in cultured medium with monocytes shows greater similarity. This occurs due to the consideration of inflammatory cells (which contain monocytes or macrophages) in the production of f-cytokines (see Eq. (\ref{M19})). Both curves indicate that the synthesis or concentration of f-cytokines increases gradually over time.
\subsection{\textbf{Comparison with Previous Models, Analyses and Limitations}}
\noindent
The model presented herein considers the multiphase structure of atherosclerotic plaque \mbox{\citep{watson2018two,watson2020multiphase}}, incorporating both mass and momentum balance approaches based on fluid mechanics. A chemotaxis function linked to the movement of inflammatory cells is included in the stress term related to the inflammatory tissue. In contrast to \mbox{\citet{watson2018two,watson2020multiphase}}, this paper argues that an early atherosclerosis model is a clear example of free boundary models as considered in several other models \mbox{\citep{fok2012growth}}, \mbox{\citep{ahmed2022macrophage,ahmed2024hdl}}. This free boundary formulation enables the direct quantification of plaque growth through the spatial domain size. The present model incorporates the influx of monocytes/macrophages into the intima through boundary condition \mbox{\eqref{SN9}}, which specifies that the influx is regulated by the concentrations of f-cytokines (MCP-1). This approach closely aligns with the biological process of atherosclerosis \citep{hansson2006immune,chalmers2015bifurcation}, rather than assuming that monocyte recruitment is solely driven by the quantity of modified LDL \mbox{\citep{ahmed2022macrophage}}. Including species, f-cytokines, enables us to model monocyte recruitment in response to f-cytokines, which ultimately play a significant role in early plaque development. Note that the flux of f-cytokines through the endothelium is influenced by oxLDL, which has been considered. Despite the critical role of f-cytokines in monocyte migration in early atherosclerosis, it is largely overlooked in spatial-free boundary models, particularly in existing multiphase models that consider chemotactic movement toward modified LDL.\\

\noindent
Intracellular cholesterol and its toxicity-induced cell death is an interesting aspect of early plaque development. To the best of our knowledge, the work by \mbox{\citet{ahmed2024hdl}} is the only study to consider the toxic effects of intracellular cholesterol. Their model assumes that cytotoxicity-induced cell death is linearly dependent on the intracellular lipid fraction, implying that toxicity-induced cell death occurs at any concentrations of intracellular cholesterol. However, cell physiology demands a reference intracellular cholesterol concentration beyond which such death may occur. This aspect is supported by in vitro evidence from \mbox{\citet{liu2008mulberry}}, who studied the effects of mulberry extracts on LDL oxidation and foam cell formation. In addition, \mbox{\citet{yang2011mulberry}} further support this finding, emphasizing that high cholesterol can induce toxicity leading to cell death. Consequently, based on this evidence, we have considered a critical threshold concentration of intracellular cholesterol $c_2^{*}$, beyond which cells experience toxicity-induced death. Present study finds that this incident resembles with the cell-cell interaction phenomena among the inflammatory cells introduced in Eq. (\mbox{\ref{M8a}}). Consequently, beyond the threshold $c_2^{*}$, the impact of toxicity or the rate of toxicity-induced cell death increases with intracellular cholesterol concentration ($c_2$). \\

\noindent
The analysis of parameters associated with intracellular cholesterol-induced toxicity reveals that it slows plaque growth, thereby contributing to stabilization. This phenomenon occurs due to an increase in the death of inflammatory cells. These findings are consistent with the medical review article by \mbox{\citet{kockx2000apoptosis}}, which is based on experimental studies. The review postulates that the loss of macrophages can lead to plaque stabilization, and if apoptotic bodies are efficiently cleared (e.g., via efferocytosis), cell death may further promote stabilization. Additionally, an experimental study by \mbox{\citet{ball1995evidence}} supports the hypothesis that macrophage foam cell death, caused by oxLDL-induced toxicity, may reduce plaque growth through increased cell death. Furthermore, our model indicates that the population of inflammatory cells is higher near the endothelium or EII. This observation is consistent with findings from histological studies on early human aortic fatty streaks \mbox{\citep{guyton1993transitional}}, which reveal that macrophage-derived foam cells are predominantly located in the middle and closer to the endothelium. Advanced plaques in humans and mice exhibit a similar structural pattern, where macrophages tend to localize nearer to the endothelium and necrotic cores form deeper within the plaque \mbox{\citep{macneill2004focal}}, \mbox{\citep{dickhout2011induction}}.\\

\noindent
The present study does not account for blood flow but begins with the assumption that the endothelium is already injured, thereby permitting LDL to penetrate the blood vessel wall. A promising extension of this work is coupling blood flow in the lumen with events in the intima to model early plaque development with a free boundary. It can be achieved by explicitly incorporating blood flow and LDL transport within the bloodstream and the subsequent influx into the intima \citep{filipovic2013computer,bulelzai2012long}. The present study simplifies several complex mechanisms, providing a good foundational framework for developing more detailed models of plaque growth in early atherosclerosis. A prominent extension is incorporating the impact of HDL, which possesses various atheroprotective properties \mbox{\citep{barter2004antiinflammatory}}, such as accepting cholesterol from foam cells and preventing LDL oxidation. Another extension could incorporate several additional factors, such as the influence of dead materials \mbox{\citep{kockx2000apoptosis}} and T cells \mbox{\citep{robertson2006t}}, as well as processes like efferocytosis. However, a limitation of our continuous multiphase modelling approach is that it does not directly account for cell numbers, as the multiphase framework is a volume-averaged continuum model. This limitation may be addressed by adopting discrete representations for early plaque processes. For example, discrete agent-based models, which are commonly used to model tissue growth \mbox{\citep{van2015simulating}}, could be applied.
\section{Concluding Remarks}
\noindent
This study adopts a biphasic mixture model to investigate the role of macrophage and foam cells (referred to as inflammatory cells) in early atherosclerosis. The role of f-cytokines in their recruitment and activation towards the inflammatory process is discussed thoroughly. OxLDL cholesterol particles get internalized through the inflammatory cells as intracellular cholesterol becomes cytotoxic to the inflammatory cells. This study identifies three regions corresponding to early plaque growth: early state, advanced state and persistent state of risk. The growth is initially steep but the growth rate gradually slows down to an approximately constant due to the toxicity effect induced by intracellular cholesterol.\\

\noindent
Additionally, this toxic effect affects the functioning of inflammatory cells. Correspondingly, a correlation \mbox{(\ref{RD1})} is proposed that dictates the plaque growth process when the cholesterol-induced toxicity issue is considered. This study introduces three influx parameters: $\nu_1$, $\nu_2$, and $\nu_3$. The first parameter, $\nu_1$, is associated with the influx of oxLDL cholesterol, while the second and third parameters, $\nu_2$ and $\nu_3$, correspond to the influx of f-cytokines. A higher influx of oxLDL cholesterol leads to a more reduced growth rate of plaque width. The influx parameters $\nu_2$ and $\nu_3$ behave oppositely to the oxLDL cholesterol influx parameter during plaque development. Both these parameters are responsible for the growth of the inflammatory cell phase and, thus, plaque width. The present analysis indicates that the oxLDL cholesterol influx parameter positively influences the flattening of the plaque growth rate.\\

\noindent
On the other hand, the influx parameters associated with f-cytokines destabilize the flattening. In addition, f-cytokines can reduce the overall inflammatory process by decreasing the recruitment of inflammatory cells from the bloodstream to the intima layer through the damaged endothelium. In addition, the parameters $s_5$ and $s_6$, linked to cholesterol toxicity-induced cell death, play a crucial role in stabilizing plaque growth. An increase in $s_5$ or a decrease in $s_6$ reduces the plaque growth rate, flattening the plaque growth profile. Hence, intracellular cholesterol-induced toxicity can slow plaque growth, resulting in a flattened growth pattern. Moreover, we observed a significant change in the plaque growth profile due to the variation in the toxicity level, $c^{*}_2$. A higher toxicity level primarily indicates a reduction in the death of inflammatory cells up to that level and the proliferation of more inflammatory cells, thereby accelerating plaque growth. Therefore, this study may contribute to developing novel therapeutic strategies to prevent atherosclerosis progression during its early stages.
\appendix
\subsection*{\textbf{Appendix A: Leading order coefficients}}\label{Appendix A}
\noindent
The expressions of ${\left(\varphi^n\right)}^{(0)}$, ${\daleth^{f}}^{(0)}$, $Q^{(0)}_1$, $Q^{(0)}_2$, $Q^{(0)}_3$ $\Sigma^{(0)}$, and $\Lambda^{(0)}$ mentioned in equations \mbox{\eqref{P4}}-\mbox{\eqref{P9}} are as
\begin{flalign*}
{\left(\varphi^{n}\right)}^{(0)}  = 1 - {\left(\varphi^{f}\right)}^{(0)}, &&
\end{flalign*}
\begin{flalign*}
{\daleth^{f}}^{(0)} & = {\left(\varphi^{f}\right)}^{(0)} {\left(\varphi^{n}\right)}^{(0)} \frac{s_0 c_3^{(0)}}{1+s_1 c_3^{(0)}} + {\left(\varphi^{f}\right)}^{(0)} {\left(\varphi^{n}\right)}^{(0)} \frac{s_2 c_1^{(0)}}{1+s_3 c_1^{(0)}}
- {\left(\varphi^{f}\right)}^{(0)} \frac{s_4 + s_5 c_2^{(0)}}{1+s_6 c_2^{(0)}}\mathcal{H}(c_2^{(0)}-c_2^*) \\
& - {\left(\varphi^{f}\right)}^{(0)}s_7,
\end{flalign*}

\begin{flalign*}
 Q_1^{(0)} = {\left(\varphi^{f}\right)}^{(0)} {\left(\varphi^{n}\right)}^{(0)} q_0 c_1^{(0)} , &&
\end{flalign*}
\begin{flalign*}
 Q_2^{(0)} = - {\left(\varphi^{f}\right)}^{(0)} {\left(\varphi^{n}\right)}^{(0)} q_0 c_1^{(0)} + {\left(\varphi^{f}\right)}^{(0)} {\left(\varphi^{n}\right)}^{(0)} q_2 c_2^{(0)} , &&
\end{flalign*}
\begin{flalign*}
 Q_3^{(0)} = {\left(\varphi^{n}\right)}^{(0)} q_3 c_3^{(0)}  - {\left(\varphi^{n}\right)}^{(0)} {\left(\varphi^{f}\right)}^{(0)} q_4 c_1^{(0)} c_3^{(0)} , &&
\end{flalign*}
\begin{flalign*}
\Sigma^{(0)} = \frac{\left({\left(\varphi^{f}\right)}^{(0)} - \varphi^*\right)}{\left( 1-{\left(\varphi^{f}\right)}^{(0)} \right)^2} \mathcal{H}({\left(\varphi^{f}\right)}^{(0)} - \varphi^*) , &&
\end{flalign*}
\begin{flalign*}
\Lambda^{(0)} = \frac{\chi}{1+ \left(\kappa c_3^{(0)}\right)^m}. &&
\end{flalign*}
\appendix
\subsection*{\textbf{Appendix B: $\mathcal{O}(q_1)$ coefficients}} \label{B}
\noindent
The expressions of ${\left(\varphi^n\right)}^{(1)}$, ${\daleth^{f}}^{(1)}$, $Q^{(1)}_1$, $Q^{(1)}_2$, $Q^{(1)}_3$ $\Sigma^{(1)}$, and $\Lambda^{(1)}$ mentioned in equations \mbox{\eqref{P4}}-\mbox{\eqref{P9}} are as
\begin{flalign*}
{\left(\varphi^{n}\right)}^{(1)} = - {\left(\varphi^{f}\right)}^{(1)}, &&
\end{flalign*}
\begin{flalign*}
{\daleth^{f}}^{(1)} & = \left({\left(\varphi^{f}\right)}^{(0)} {\left(\varphi^{n}\right)}^{(1)} + {\left(\varphi^{f}\right)}^{(1)} {\left(\varphi^{n}\right)}^{(0)}\right) \frac{s_0 c_3^{(0)}}{1+s_1 c_3^{(0)}} + {\left(\varphi^{f}\right)}^{(0)} {\left(\varphi^{n}\right)}^{(0)} \left(\frac{s_0 c_3^{(1)}}{1+s_1 c_3^{(0)}} - \frac{s_0 s_1 c_3^{(0)}c_3^{(1)}}{\left(1+s_1 c_3^{(0)}\right)^2}\right) \\
                 & + \left({\left(\varphi^{f}\right)}^{(0)} {\left(\varphi^{n}\right)}^{(1)} + {\left(\varphi^{f}\right)}^{(1)} {\varphi^{n}}^{(0)}\right) \frac{s_2 c_1^{(0)}}{1+s_3 c_1^{(0)}} + {\left(\varphi^{f}\right)}^{(0)} {\left(\varphi^{n}\right)}^{(0)} \left(\frac{s_2 c_1^{(1)}}{1+s_3 c_1^{(0)}} - \frac{s_2 s_3 c_1^{(0)}c_1^{(1)}}{\left(1+s_3 c_1^{(0)}\right)^2}\right) \\
                 & - \left( {\left(\varphi^{f}\right)}^{(1)} \frac{s_4 + s_5 c_2^{(0)}}{1 + s_6 c_2^{(0)}} + {\left(\varphi^{f}\right)}^{(0)} \left( \frac{s_5 c_2^{(1)}}{1+s_6 c_2^{(0)}} - \frac{s_6 c_2^{(1)}(s_4 + s_5 c_2^{(0)})}{\left(1+s_6 c_2^{(0)}\right)^2} \right) \right) \mathcal{H}(c_2^{(1)} - c_2^*)- {\left(\varphi^{f}\right)}^{(1)} s_7,
\end{flalign*}
\begin{flalign*}
 Q_1^{(1)} = {\left(\varphi^{f}\right)}^{(0)} {\left(\varphi^{n}\right)}^{(0)} q_0 \left(c_1^{(1)} - (c_1^{(0)})^2\right) + \left({\left(\varphi^{f}\right)}^{(0)}{\left(\varphi^{n}\right)}^{(1)} + {\left(\varphi^{f}\right)}^{(1)}{\left(\varphi^{n}\right)}^{(0)} \right) q_0 c_1^{(0)}, &&
\end{flalign*}
\begin{flalign*}
 Q_2^{(1)}  &  =  - {\left(\varphi^{f}\right)}^{(0)} {\left(\varphi^{n}\right)}^{(0)} q_0 \left(c_1^{(1)} - (c_1^{(0)})^2\right) - \left({\left(\varphi^{f}\right)}^{(0)}{\left(\varphi^{n}\right)}^{(1)} + {\left(\varphi^{f}\right)}^{(1)}{\left(\varphi^{n}\right)}^{(0)} \right) q_0 c_1^{(0)}  \\
   & + {\left(\varphi^{f}\right)}^{(0)}{\left(\varphi^{n}\right)}^{(0)}q_2 c_2^{(1)} +  \left({\left(\varphi^{f}\right)}^{(0)}{\left(\varphi^{n}\right)}^{(1)} + {\left(\varphi^{f}\right)}^{(1)}{\left(\varphi^{n}\right)}^{(0)} \right) q_2 c_2^{(0)},
\end{flalign*}
\begin{flalign*}
 Q_3^{(1)} & = q_3 \left({\left(\varphi^{n}\right)}^{(0)} c_3^{(1)} + {\left(\varphi^{n}\right)}^{(1)} c_3^{(0)} \right) - q_4 \left({\left(\varphi^{n}\right)}^{(0)} {\left(\varphi^{f}\right)}^{(1)} + {\left(\varphi^{n}\right)}^{(1)} {\left(\varphi^{f}\right)}^{(0)} \right) c_1^{(0)} c_3^{(0)}\\
& - q_4 {\left(\varphi^{n}\right)}^{(0)} {\left(\varphi^{f}\right)}^{(0)} \left(c_1^{(0)} c_3^{(1)} + c_1^{(1)} c_3^{(0)}\right),
\end{flalign*}
\begin{flalign*}
\Sigma^{(1)} = \left( \frac{{\left(\varphi^{f}\right)}^{(1)}}{\left(1-{\left(\varphi^{f}\right)}^{(0)} \right)^2} + \frac{2 ({\left(\varphi^{f}\right)}^{(0)} - \varphi^*) {\left(\varphi^{f}\right)}^{(1)}}{\left(1-{\left(\varphi^{f}\right)}^{(0)} \right)^3}\right) \mathcal{H}({\left(\varphi^{f}\right)}^{(1)} - \varphi^*), &&
\end{flalign*}
\begin{flalign*}
\Lambda^{(1)} = -\frac{m \chi \kappa^m \left( c_3^{(0)}\right)^{m-1} c_3^{(1)}}{\left(1+\kappa^m \left( c_3^{(0)}\right)^{m}\right)^2}. &&
\end{flalign*}


\end{document}